\documentclass[12pt]{article}
\usepackage[utf8]{inputenc}
\usepackage[english]{babel}
\usepackage{amsthm}

\usepackage{epsfig}
\usepackage[dvipdfmx]{color}
\usepackage{amsmath}
\usepackage{amssymb}
\usepackage{lscape}
\usepackage{bm}

\makeatletter
\@addtoreset{equation}{section}

\makeatother

\begin{document}

\title{
\vbox{
\baselineskip 14pt
\hfill \hbox{\normalsize KEK-TH-2093
}} \vskip 1cm
\bf \Large  Effective Potential for Revolving D-branes
\vskip 0.5cm
}
\author{
Satoshi Iso$^{a,b}$  \thanks{E-mail: \tt iso(at)post.kek.jp}, 
Hikaru Ohta$^{a,b}$ \thanks{E-mail: \tt hohta(at)post.kek.jp}, 
Takao Suyama$^{a}$ \thanks{E-mail: \tt tsuyama(at)post.kek.jp}
\bigskip\\
\it \normalsize
$^a$ Theory Center, High Energy Accelerator Research Organization (KEK), \\
\it  \normalsize 
$^b$Graduate University for Advanced Studies (SOKENDAI),\\
\\
\it Tsukuba, Ibaraki 305-0801, Japan \\
\smallskip
}
\date{\today}

\maketitle

\abstract{\normalsize
We quantize an open string stretched between D0-branes revolving around each other.
The worldsheet theory  is analyzed in a rotating coordinate system 
in which the worldsheet fields obey simple boundary conditions, but instead
the worldsheet Lagrangian becomes nonlinear.
We quantize the system perturbatively with respect to the velocity of the D-branes
and determine the one-loop partition function of the open string, 
from which we extract the short-distance behavior of the effective potential for the revolving D0-branes. 
It is compared with the calculation of the partition function of open strings 
between D0-branes moving at a constant relative velocity. 

\newpage

\section{Introduction}

D-branes in the superstring theory have played pivotal roles  in understanding  nonperturbative behaviors 
of string theory. They are also widely used in the string phenomenology and cosmology (for review see
\cite{Blumenhagen:2006ci,Ibanez:2012zz,Baumann:2014nda} and references therein). 
In many cases,  static configurations of D-branes  are considered, 
especially those with a fraction of supersymmetry preserved. 
It is partially because these configurations are of particular interest in mathematical settings
but also because the analysis is simple and exact calculations can be performed.
However, in many  phenomenologically or cosmologically
 interesting situations,
D-branes are moving and no supersymmetries are preserved. 
For example, if our universe is described by the brane-world scenario
\cite{Dvali:1998pa,Kehagias:1999vr,Silverstein:2003hf,Easson:2007dh}, 
those branes  may have experienced irregular motions in the very early universe.
In particular, if D-branes are accelerating to each other, 
they would emit closed string radiation \cite{AbouZeid:1999fs,Mironov:2007nk,Bachlechner:2013fja}, 
and particle creation of open strings would occur \cite{Bachas:1995kx}.  
Then we may ask \cite{Kofman:2004yc}\cite{Enomoto:2013mla}: 
What are the final configurations of such moving D-branes?  
Do D-branes collapse or scatter away from each other? 
In the D-brane scenario of universe and in the D-brane constructions of 
the standard model of particle physics, 
answering these questions will be relevant to study stability of our universe 
as well as the moduli stabilization, which may include the hierarchy problem 
of the electroweak scale against various UV scales.

As a first step towards answering these questions, 
we study a pair of D0-branes of bosonic string theory
which revolve around each other in the flat space-time and calculate potential between them.
In this paper, we will not discuss the underlying mechanism of the rotation, but instead, 
we analyze properties of open strings stretched between such a pair of D0-branes\footnote{
This is certainly different from a freely rotating classical open string which can be readily analyzed. 
}.  
In particular, potential between D0-branes is read from the one-loop partition function 
of the open strings. 
If D0-branes are far from each other, the system is more appropriately described by the 
closed strings and the potential is given by the gravitational potential. 
In \cite{Kazama:1997bc}, the amplitude for the exchange of a single closed string between two D0-branes was obtained. 
The result is for D0-branes moving along arbitrary trajectories with small accelerations, and includes contributions from all massive closed string modes. 
See also \cite{Hirano:1996pf}. 
On the other hand, if the distance is shorter than
the string scale, massless open string modes dominate and the effective dynamics 
of D0-branes are described by the DBI action, or the Yang-Mills action if we neglect higher derivative terms
\cite{Bachas:1995kx}\cite{Douglas:1996yp}\cite{Lifschytz:1996iq}. 
In this paper, in order to take into account  massive open string states as well as the massless states, 
we calculate one-loop partition function of open strings stretched 
between revolving D0-branes. 
We are interested in  the behavior when 
the relative distance of D0-branes is shorter than the string length \cite{Douglas:1996yp}. 

Let us here recall the well-known result of the one-loop open string partition function $Z=-{\cal V} {\cal T}$
between D0-branes at rest with relative distance $y$. It is given \cite{Polchinski:1998rq} by 
\begin{equation}
Z = \int_0^\infty \frac{ds}{2s} {\rm Tr} \left[ e^{-2\pi s L_0} \right] 
= {\cal T} \int_0^\infty \frac{ds}{2s} (8\pi^2\alpha' s)^{-\frac12} e^{-\frac{y^2}{2\pi \alpha'}s} \eta(is)^{-24}
\label{partition-function-atrest}
\end{equation}
where $\eta(is)=e^{-2\pi s/24} \prod_{m=1}^\infty (1-e^{-2\pi ms})$ and ${\cal T}$ is the time 
duration of the configuration.
If $y \gg \sqrt{\alpha'}$, we can use the modular transformation $\eta(is)=s^{-1/2} \eta(i/s)$ and the expansion
$\eta(i/s)^{-24}=\exp(2\pi /s)+24+ \cdots$ to obtain the effective potential ${\cal V}(y) \propto 24/y^{23}$
that is dominated by the closed string massless modes such as a graviton, if the closed string tachyon
is neglected.
In the present paper, we are interested in the opposite limit $y \ll \sqrt{\alpha'}$
where low energy open string modes dominate the potential.  
The Dedekind $\eta$-function can be expanded as
$ \eta(is)^{-24} =   \sum_{n=-1}^\infty c_{n} e^{-2 n \pi s} $
 where $c_{-1}=1, c_0=24, c_1=324, c_2=3200$ and so on.
Then the effective potential ${\cal V}$ can be calculated as a sum
\begin{eqnarray}
{\cal V}(y) &=& -\frac{1}{\sqrt{8\pi^2 \alpha'}} \int_0^\infty \frac{ds}{2s^{3/2}} 
\sum_{n=0}^\infty c_{n} e^{-\left( 2 n \pi+\frac{y^2}{2\pi \alpha'} \right) s}
 \nonumber \\
 &=& -\sum_{n=0}^\infty \frac{c_{n} }{ \sqrt{8 \pi^2 \alpha'}} \sqrt{ 2\pi n+ \frac{y^2}{2\pi \alpha'}}
 \int_0^\infty \frac{ds}{2s} s^{-\frac12} e^{-s}.
 \label{EPatrest}
\end{eqnarray}
where the open string tachyon ($c_{-1}$) is neglected.
The $s$-integral is UV divergent at $s=0$. 
Here we simply evaluate it by an analytical continuation, which gives 
a constant (see (\ref{integral})). 
As a function of the distance $y$,
the effective potential ${\cal V}(y)$ behaves regularly near the origin $y=0$.
The massless contribution $n=0$ gives a linear potential $r$ while the
massive contributions can be
expanded in positive powers of $y^2$, and we get the potential
${\cal V}=\sum_{n=0}^\infty {\cal V}_n$;
\begin{eqnarray}
{\cal V}_0 (y) &=&  \frac{12}{\sqrt{\alpha'}}  \sqrt{ \frac{y^2}{4 \pi^2 \alpha'} }
\label{V1atrest}
\\
 {\cal V}_1(y) &=& \frac{162}{\sqrt{\alpha'}}  
 \left( 
 1 + \frac{1}{2} \left( \frac{y^2}{4\pi^2 \alpha'} \right) - \frac{1}{8}  \left( \frac{y^2}{4\pi^2 \alpha'} \right) ^2 
 + \cdots
 \right) .
 \label{V2atrest}
\end{eqnarray}
Here ${\cal V}_0(y)$ is the massless mode contribution and ${\cal V}_1(y)$ is the first excited massive mode contribution. 
What we would like to study in the present paper
 is the behavior of the effective potential ${\cal V}(y, \omega)$ near $y=0$
 when D0-branes are  revolving around each other with angular velocity $\omega$ and distance $y=2r$. 
Since the potential is an analytic function of the velocity $v=\omega r$ 
  of each D0-brane, it is naturally expected that the mass squared
$m_n^2 \equiv (n+ y^2/(2\pi)^2 \alpha')$ is replaced by 
something like 
$(n+f_i(r^2/\alpha', v))$ where $f_i(r^2/\alpha', v)$ is a regular function of $r^2$ and $v$, and 
$i$ specifies a different excitation.
The function $f_i$ is, of course, reduced to  $f_i(r^2/\alpha', 0)=(2r)^2/(2\pi)^2 \alpha'$ 
when D0-branes are at rest\footnote{ The partition function is invariant under $v \rightarrow -v$ and
an even function of the velocity $v$. But it does not mean that the energy eigenvalue is a function of $v^2$.
Rather, a pair of eigenvalues appear that are exchanged under the symmetry $v \rightarrow -v$ as seen
in the discussion after Eq. (\ref{exact 1-loop}).}. 
For this purpose, we perturbatively calculate the partition function of open strings 
stretched between such revolving D0-branes, which is a generalization of Eq. (\ref{partition-function-atrest})
and given in Eq. (\ref{partition function final}). For comparison, we also calculate the partition function of open strings
stretched between D0-branes moving at a constant relative velocity $2v$, which is given in Eq. (\ref{exact 1-loop}). 
Eq. (\ref{partition function final}) is the main result of the present paper. 

The action of the worldsheet theory is simply given by
\begin{equation}
S=-\frac{1}{4\pi \alpha'}\int d^2\sigma\, \partial_\alpha X_{\mu}\partial^\alpha X^{\mu}. 
   \label{non-rotating action}
\end{equation} 
The quantization of this theory is, however, not so simple 
since the boundary conditions of $X^\mu$ are complicated due to the rotation
of D0-branes at the ends of the open strings.  
Actually, 
since the D0-branes are moving, the boundary conditions depend on the value of $X^0$
and the spacial and temporal coordinates are mixed with other. 
To overcome this difficulty,
we study the worldsheet theory by employing the rotating coordinate system in the target space. 
In this formulation, the boundary conditions of the worldsheet fields become simple, 
but instead,  the coordinate transformation generates
higher order terms of the world sheet fields in the action  and makes the system nonlinear.
We thus analyze this system perturbatively with respect to the velocity of the D0-branes
and calculate the one-loop partition function of the open strings, from which
the short distance behavior of the potential between the revolving D0-branes is extracted.

This paper is organized as follows. 
In section \ref{worldsheet}, we quantize open strings stretched between revolving D0-branes.
We reformulate the worldsheet theory of the open strings 
so that the boundary conditions of the worldsheet fields become simple,
but in compensation, the system becomes interacting
and the resulting theory must be quantized perturbatively. 
In section \ref{partition function}, we calculate the one-loop partition function, based on the formalism developed in 
\cite{Iso:2017hvi}. 
This is regarded as the effective potential for the D0-branes.  
The short-distance behavior is investigated in section \ref{effective potential}. 
In section \ref{linear}, 
we make a comparison of our result with the one obtained from D0-branes moving with constant velocities. 
Section \ref{discussion} is devoted to conclusions and discussions. 
Details of the calculations are summarized in Appendices.

\section{Open strings stretched between revolving D-branes
} \label{worldsheet}

\subsection{Open strings in the rotational coordinate system}
As a simple system of rotating D-branes in bosonic string, 
we consider two D0-branes in the flat space-time  revolving around each other like a binary star. 
We assume that the orbits of the D0-branes lie on the $x$-$y$ plane. 
Their positions $(x_1,y_1)$ and $(x_2,y_2)$  are changing with time and given by 
\begin{equation}
\left\{
\begin{array}{l}
x_1(t)\ =\ r\cos\omega t, \\ [2mm]
y_1(t)\ =\ r\sin\omega t, 
\end{array}
\right. \hspace{1cm} \left\{
\begin{array}{l}
x_2(t)\ =\ -r\cos\omega t, \\ [2mm]
y_2(t)\ =\ -r\sin\omega t. 
\end{array}
\right.
   \label{orbit}
\end{equation}
The revolving motion of the D0-branes is not a classical solution unless
there is an attractive potential between D-branes.
Here we implicitly assume that it is generated  either by exchanges of closed strings 
between D0-branes or by introducing suitable background fields. 
In the following we restrict ourselves to consider 
the situation in which the angular frequency of the D0-branes is small so that
the background fields can be treated perturbatively around the flat background space-time. 
Analysis including the backreaction is left for future investigations. 

We then consider an open string stretched between these D0-branes. 
As explained in the introduction, we choose the rotating coordinate system for the target space-time in which the D0-branes are static. 
In the original coordinate system used in the action (\ref{non-rotating action}), the target space metric is 
simply given by 
\begin{equation}
ds^2 = -dt^2+dx^2+dy^2+(dx^i)^2
\end{equation}
where $i=3,4,\cdots,25$. 
We then introduce the rotating coordinate system  defined by
\begin{eqnarray}
\tilde{t} &:=& t, \\
\tilde{x} &:=& x\cos\omega t+y\sin\omega t, \\
\tilde{y} &:=& -x\sin\omega t+y\cos\omega t, \\
\tilde{x}^i &:=& x^i. 
\end{eqnarray}
In this coordinate system, the orbits  of the D0-branes (\ref{orbit}) become static 
\begin{equation}
\left\{
\begin{array}{l}
\tilde x_1(t)\ =\ r, \\ [2mm]
\tilde y_1(t)\ =\ 0, 
\end{array}
\right. \hspace{1cm} \left\{
\begin{array}{l}
\tilde x_2(t)\ =\ -r, \\ [2mm]
\tilde y_2(t)\ =\ 0 
\end{array}
\right. 
\end{equation}
but the metric takes the following non-diagonal form:
\begin{equation}
ds^2 = -d\tilde{t}^2+d\tilde{x}^2+d\tilde{y}^2+2\omega d\tilde{t}(\tilde{x}d\tilde{y}-\tilde{y}d\tilde{x})+\omega^2(\tilde{x}^2+\tilde{y}^2)d\tilde{t}^2+(d\tilde x^i)^2. 
\end{equation}
Accordingly, the worldsheet action becomes 
\begin{eqnarray}
S 
&=& -\frac{1}{4\pi\alpha'}\int d^2\sigma\left[ -\partial_\alpha \tilde{T}\partial^\alpha \tilde{T}+\partial_\alpha \tilde{X}\partial^\alpha \tilde{X}+\partial_\alpha \tilde{Y}\partial^\alpha \tilde{Y}+\partial_\alpha \tilde{X}^i\partial^\alpha \tilde{X}_i \right. \nonumber \\ [1mm]
& & \left. \hspace{1cm}+2 \omega \partial_\alpha \tilde{T}(\tilde{X}\partial^\alpha \tilde{Y}-\tilde{Y}\partial^\alpha \tilde{X})+
\omega^2(\tilde{X}^2+\tilde{Y}^2)\partial_\alpha \tilde{T}\partial^\alpha \tilde{T} \right] .
   \label{action-before}
\end{eqnarray}
We have chosen the conformal gauge. 
This is allowed if the D0 particles satisfy the equation of motion and 
the conformal symmetry is preserved. 
In the present case, it is slightly violated at the boundaries, but the violation is 
expected to be small as far as the angular frequency is small compared to the string scale. 
The modification of the target space metric due to the external field is treated as vertex operator insertions.
In the following calculations, we ignore such corrections to the effective potential
under an assumption   $\omega/m_{\text str} < 1$.
 We want to come back to this issue in future publications. 

By rescaling the fields as $\tilde{X}^\mu\to r\tilde{X}^\mu$, the action is given by
\begin{eqnarray}
S 
&=& -\frac{r^2}{4\pi\alpha'}\int d^2\sigma\left[ -\partial_\alpha \tilde{T}\partial^\alpha \tilde{T}+\partial_\alpha \tilde{X}\partial^\alpha \tilde{X}+\partial_\alpha \tilde{Y}\partial^\alpha \tilde{Y}+\partial_\alpha \tilde{X}^i\partial^\alpha \tilde{X}_i \right. \nonumber \\ [1mm]
& & \left. \hspace{1cm}+2v\partial_\alpha \tilde{T}(\tilde{X}\partial^\alpha \tilde{Y}-\tilde{Y}\partial^\alpha \tilde{X})+v^2(\tilde{X}^2+\tilde{Y}^2)\partial_\alpha \tilde{T}\partial^\alpha \tilde{T} \right], 
   \label{action}
\end{eqnarray}
where $v:=r\omega$ is the velocity of the 
D0-branes.\footnote{Note that the relative velocity is $2v$ and the distance is $y=2r$.}  
The boundary conditions for the rescaled fields, $\tilde{X}$ and $\tilde{Y}$, are simple
in the  new coordinate system, 
\begin{equation}
\tilde{X}(\tau,\sigma)\ =\ \left\{
\begin{array}{cc}
+1, & (\sigma=0) \\ [2mm]
-1, & (\sigma=\pi)
\end{array}
\right. \hspace{1cm} 
\tilde{Y}(\tau,\sigma)\ =\ \left\{
\begin{array}{cc}
0, & (\sigma=0) \\ [2mm]
0. & (\sigma=\pi)  
\end{array}
\right. 
   \label{b.c. for X and Y}
\end{equation}
However, the boundary condition for $\tilde{T}$ is still nontrivial. 
Indeed, the variation of the action with respect to $\tilde{T}$  gives the following boundary term:
\begin{eqnarray}
\delta S \Big|_{bdy}
  =  -\frac{r^2}{2\pi\alpha'}  
 \delta \tilde T\left[ -\partial_\sigma\tilde T+v(\tilde X\partial_\sigma\tilde Y-\tilde Y\partial_\sigma\tilde X)+v^2(\tilde X^2+\tilde Y^2)\partial_\sigma\tilde T \right]\Big|_{bdy}  \nonumber \\
\end{eqnarray}
By using (\ref{b.c. for X and Y}), 
$\tilde T$ must satisfy 
\begin{equation}
\begin{array}{ccc}
(1-v^2)\partial_\sigma\tilde{T}-v  \partial_\sigma\tilde{Y}\ =\ 0,
 &\hspace{5mm}& (\sigma=0) \\ [2mm]
(1-v^2)\partial_\sigma\tilde{T}+v\partial_\sigma\tilde{Y}\ =\ 0. &\hspace{5mm}& (\sigma=\pi)
\label{Tboundary}
\end{array}
\end{equation}
These conditions are linear in the world sheet fields 
since we have substituted the boundary values of (\ref{b.c. for X and Y})
for the nonlinear terms that  would have appeared  in (\ref{Tboundary}). 
To simplify these conditions, we define a new field $T$ as 
\begin{equation}
T\ :=\ \tilde{T}-\frac{v}{1-v^2}x(\sigma)\tilde Y
\label{new-field-T}
\end{equation}
where 
\begin{equation}
x(\sigma)\ :=\ 1-\frac{2\sigma}{\pi}.
\end{equation}
Using the boundary condition (\ref{b.c. for X and Y}) for $\tilde Y$, it can be shown that the field $T$ satisfies the ordinary Neumann boundary condition $\partial_\sigma T|_{\sigma=0,\pi}=0$.
We also introduce a new field $X$ 
\begin{equation}
\tilde{X}\ =\ x(\sigma)+X. 
\end{equation} 
Then  $X|_{\sigma=0,\pi}= Y|_{\sigma=0,\pi}=0$ are satisfied.
For notational simplicity, we use tilde-less notations for $Y=\tilde Y$ and  $X_i=\tilde{X}_i$ 
in the following discussions.

\vspace{5mm}
To summarize,
the new world sheet fields $X,Y,T$ and $X_i$ (for $i=3, \cdots 25$)
satisfy the boundary conditions
\begin{equation}
X|_{\sigma=0,\pi}=0,\hspace{5mm} Y|_{\sigma=0,\pi}=0,\hspace{5mm}\partial_\sigma T|_{\sigma=0,\pi}=0, 
\hspace{5mm}{X_i |_{\sigma=0,\pi}=0}
   \label{simple b.c.}
\end{equation}
and the world sheet action is given by 
\begin{eqnarray}
S 
&=& -\frac{r^2}{4\pi\alpha'}\int d^2\sigma
 \Bigg[ -\dot{X}^2-\dot{Y}^2-(\dot{X}^i)^2 
+\left(  X'  - \frac{2}{\pi} \right)^2 
+(Y')^2  + ({X^i}' )^2 
 \nonumber \\
 &+& \left[1- v^2 \Big( \big(X+x(\sigma) \big)^2
+Y^2  \Big) \right]
\nonumber \\
&& \hspace{15mm} \times
\left[
\left( \dot{T} + \frac{v x(\sigma)}{1-v^2}   \dot{Y}  \right)^2 
- \left( T' + \frac{v}{1-v^2} \big( x(\sigma) Y' -\frac{2}{\pi} Y \big)
\right)^2 
\right]
\nonumber \\
&-&  2v 
\left( \dot{T} + \frac{v x(\sigma)}{1-v^2}   \dot{Y}  \right) 
\Big( \big(X+x(\sigma)\big) \dot{Y}- Y  \dot{X}  \Big) \nonumber \\
&+&  2v \left(  
T' + \frac{v}{1-v^2} \big( x(\sigma) Y' -\frac{2}{\pi} Y \big)
\right)
\Big( \big(X+x(\sigma)\big) Y'- Y  \big(X'  - \frac{2}{\pi} \big) \Big)
 \Bigg] .
 \nonumber \\
   \label{action2}
\end{eqnarray}
In the rotating coordinate system,  
the action (\ref{action2}) becomes  nonlinear in compensation for the simple boundary conditions.
We analyze this theory perturbatively in the nonrelativistic limit $v \ll 1$.

\subsection{Perturbative Hamiltonian with respect to  $v$}
The worldsheet Hamiltonian $H$ can be obtained in the standard manner. 
It is decomposed into two parts: 
\begin{equation}
H\ =\ H_{\rm rot}(X, Y, T) +H_{\rm free} (X^i), 
\end{equation}
where $H_{\rm rot}$ governs the subsystem consisting of $X,Y$ and $T$
 while $H_{\rm free}$ is a free Hamiltonian for $X^i$. 
We now focus on the non-trivial part $H_{\rm rot}$. 
Let $\Pi_X, \Pi_Y$ and $\Pi_T$ denote the canonical momenta of $X,Y$ and $T$, respectively. 
$H_{\rm rot}$ is given in perturbative series with respect to the velocity $v$ as
\begin{equation}
H_{\rm rot}=H_0+v V_1+v^2V_2+\mathcal{O}(v^4),
\label{Hrot}
\end{equation}
where 
\begin{eqnarray}
H_0
&=& \int^\pi_0d\sigma \Biggl[\frac{\pi \alpha'}{r^2}(-{\Pi}_T^2+{\Pi}_X^2+{\Pi}_Y^2)+\frac{r^2}{4\pi \alpha'}\left\{ -(\partial_\sigma{T})^2+(\partial_\sigma{X})^2+(\partial_\sigma{Y})^2 \right\}\Biggr] 
\nonumber \\
&&  \hspace{10mm} 
+\frac{r^2}{\pi^2\alpha'} , \\
   \label{H_0-classical}
V_1&=&\int^{\pi}_{0}d\sigma \Biggl[\frac{2\pi \alpha'}{r^2}{\Pi}_T( X{\Pi}_Y-{\Pi}_X Y)+\frac{r^2}{2\pi \alpha'}\partial_\sigma  T( X\partial_\sigma Y-\partial_\sigma X Y)+\frac{2r^2}{\pi^2\alpha'}\partial_\sigma TY\Biggr],  \nonumber \\ 
   \label{V_1} 
\end{eqnarray}
\begin{eqnarray}
V_2&=&\int^{\pi}_{0}d\sigma \Biggl[\frac{\pi \alpha'}{r^2}\left\{ -(X\Pi_Y-Y\Pi_X)^2-x(\sigma)^2\Bigl({\Pi}_T^2+{\Pi}_Y^2\Bigr) \right. \nonumber \\
&&\hspace{0mm} -\left. 2x(\sigma)\Bigl({\Pi}_T^2{X}+({X}{\Pi}_Y-Y{\Pi}_X ){\Pi}_Y\Bigr)
\right\}\nonumber \\
&&\hspace{0mm}+\frac{r^2}{4\pi \alpha'}\Biggl \{(\partial_\sigma T)^2(X^2+Y^2)+x(\sigma)^2\Bigl((\partial_\sigma T)^2+(\partial_\sigma Y)^2\Bigr)+\frac{4}{\pi}x(\sigma) Y\partial_\sigma Y-\frac{12}{\pi^2}Y^2\nonumber \\
&&\hspace{0mm}+2x(\sigma)\left((\partial_\sigma T)^2 X+X(\partial_\sigma Y)^2-\partial_\sigma X Y\partial_\sigma Y\right)+\frac{4}{\pi}\left(\partial_\sigma  X Y^2- X Y\partial_\sigma Y\right)\hspace{0cm}\Biggr \}\Biggr].\nonumber \\ 
   \label{V_2}
\end{eqnarray}

In the next section, we quantize the Hamiltonian up to the second orders of $v$ and calculate
the one-loop partition function of the
open string to obtain the potential between revolving D0-branes.

\section{One-loop partition function} \label{partition function}
We are interested in the one-loop partition function of the rotating open string. 
It is given by
\begin{eqnarray}
Z&=&\int_0^\infty\frac{ds}{2s}\, \mathrm{Tr}\bigl[e^{-2\pi s (H_{\rm rot}-\frac18)}\bigr]\bigl(\eta(is)\bigr)^{-21}
\nonumber \\
&=&\int_0^\infty\frac{ds}{2s}\, \mathrm{Tr}\bigl[e^{-2\pi s (H_{\rm rot}-1)}\bigr] \prod_{m=1}^{\infty}(1-e^{-2\pi ms})^{-21}
\label{F}
\end{eqnarray}
where  $\eta(is) =e^{-2 \pi s/24} \prod_{m=1}^{\infty} (1-e^{-2\pi ms }) $.
The contributions from the $X_i$ fields in  $H_{\text free}$ ($i=3, \cdots, 25)$
 and the $(b,c)$-ghosts 
have been included in this expression. 
The one-loop determinant of a scalar field is written as
\begin{equation}
\det (\Delta + m^2)^{-1/2} = \exp \left[   \int \frac{ds}{2s} {\text Tr} e^{-(\Delta + m^2)s} \right] 
\equiv e^{-{\cal V}{\cal T}},
\end{equation}
where ${\cal T}$ is the time duration. 
Thus the open string one-loop partition function (\ref{F})
gives the minus of an effective potential ${\cal V}(r,\omega)$,
 integrated over time\footnote {\label{footnote-0mode} The stationary system of revolution is invariant under
 the time translation and the effective potential is 
 given by removing the zero mode integral of $T$. 
 When we compare the calculation with the constant velocity system in Sec.\ref{linear} 
 the integration needs  a care.},
for the two D0-branes: $Z= -{\cal V}{\cal T}.$
By the open-closed string duality, it is written as
 an exchange of a single closed string and approximately described by 
low energy closed string modes when the distance $2r$ is much larger than the string length.
In section \ref{effective potential}, we will instead
investigate the short distance behavior of the effective potential ${\cal V}(r,\omega)$
at which open string low energy modes dominate. 

In this section, we will calculate the partition function (\ref{F}) perturbatively with respect to $v$. 
In the following, we perform the calculation in the Euclidean space by the Wick rotation
of the world sheet variable from $T\to -iT$.  
Thus the velocity $v$ is also analytically continued to $iv$.

\subsection{Improved perturbation}
In order to calculate the partition function, we need to know  the energy spectrum
of the Hamiltonian. We will perform  perturbative calculations with respect to $v$, 
but even perturbatively it  is not so straightforward since
$H_0,V_1,V_2$ are not commutable with each other and we may need to diagonalize the Hamiltonian. 
Instead of explicitly diagonalizing it, we use 
a  method of the improved perturbation \cite{Iso:2017hvi}. 
This method is briefly reviewed in Appendix \ref{IOS}.  
The basic idea is that we can systematically construct a new Hamiltonian 
$H_0(v) = \sum_{n=0}^{\infty} v^n H_n$ such that 
it has the same eigenvalues as those of the original Hamiltonian 
$H=H_{\rm rot}$, but
each term $H_n$ commutes with $H_0$. 
Once $H_n$ is explicitly given, it is straightforward to take a trace to obtain the partition function.

In the following, we calculate up to the second order of the $v$-expansion:
\begin{equation}
H_0(v) := H_0+vH_1+v^2H_2+\mathcal{O}(v^4) .
   \label{H_0(v)}
\end{equation}
The first term $H_0$ is given by the original free Hamiltonian in (\ref{H_0-classical}). 
As mentioned above, $H_0(v)$ shares the same eigenvalues 
with $H_{\rm rot}$ to all orders in $v$, so we can replace $H_{\rm rot}$ with $H_0(v)$ within the perturbation theory. 
Also note that $H_1$ and $H_2$ can be constructed such that they commute with $H_0$. 
In our case, their explicit forms are given by
\begin{equation}
H_1\ :=\ V_{1,0}, \hspace{1cm} H_2:=V_{2,0}-\sum_{m\ne0}\frac1mV_{1,-m}V_{1,m} .
   \label{H_2-def}
\end{equation}
Here the operator $V_i$ $(i=1,2)$ is decomposed into $V_{i,m}$  as 
\begin{equation}
V_i = \sum_{m}V_{i,m}, \hspace{1cm} \left[ H_0,V_{i,m} \right] = mV_{i,m}
\end{equation}
where $m$'s are eigenvalues of the free Hamiltonian $H_0$.
In particular, $V_{i,0}$ is the operator contained in $V_i$ that are commutable with $H_0.$

\subsection{Perturbative calculations of the trace}
Now we calculate the trace
in the partition function (\ref{F}) with the Hamiltonian $H_{\rm rot}$ in (\ref{Hrot}).
By using the method of improved perturbation, the difficulty of calculation ${\rm Tr}e^{-2\pi s H_{\rm rot}}$
due to the non-commutativity of $H_0$ and $V_i$ is resolved by rewriting the trace
in terms of a much easier improved Hamiltonian, $H_0(v)$. 
Due to the commutativity of $H_n$ in $H_0(v)$, 
it can be expanded as
\begin{eqnarray}
 \mathrm{Tr}\bigl[e^{-2\pi s H_{\rm rot} } \bigr]  &=&
\mathrm{Tr}\bigl[e^{-2\pi s H_0(v)}\bigr]    \nonumber \\
&& \hspace{-30mm}  =
\mathrm{Tr}\bigl[e^{-2\pi s H_0}\bigr]-2\pi v s\mathrm{Tr}\bigl[e^{-2\pi s H_0}H_1\bigr] \nonumber \\ [2mm]
 && \hspace{-25mm} -
 2\pi v^2\left( s\mathrm{Tr}\bigl[e^{-2\pi s H_0}H_2\bigr]-\pi s^2\mathrm{Tr}\bigl[e^{-2\pi s H_0}H_1^2\bigr] \right) 
   \label{expand-Tr}
\end{eqnarray}
up to ${\cal O}(v^2).$
The traces are taken over the eigenstates $|\vec{\bm{n}};k\rangle$ of $H_0$
where $k$ is the momentum and $\vec{\bm{n}}$ represent the eigenmodes of harmonic oscillators. 
Since $H_0$ is a free Hamiltonian, and $X,Y$ and $T$ obey simple boundary conditions (\ref{simple b.c.}), 
the eigenvalues and eigenstates of $H_0$ are explicitly given.
In the following, we recall them for fixing the notations. 

The free Hamiltonian $H_0$ is given in terms of the mode operators as 
\begin{equation}
H_0\ =\ \alpha'p^2+\frac{r^2}{\pi^2\alpha'}+ {\tilde H}_0 
   \label{H_0}
\end{equation}
where 
\begin{equation}
{\tilde H}_0 = \sum_{\mu=T,X,Y} \sum_{n=1}^\infty N_{\mu,n}, \hspace{10mm}
N_{\mu,n} : = \ \ :\alpha^\mu_{-n}\alpha^\mu_{n}:.
\label{tildeH_0}
\end{equation}
For our conventions of the mode expansions, see Appendix \ref{free theory}. 
Note that $N_{\mu,-n}=N_{\mu,n}$ hold. 
We denote the eigenstates of $H_0$ as $|\vec{\bm{n}};k\rangle$ which is given by
\begin{equation}
|\vec{\bm{n}};k\rangle= \prod_{m_X, m_Y, m_T=1}^\infty 
(\alpha_{-m_X}^X)^{n^X_{m_X}} (\alpha_{-m_Y}^Y)^{n^Y_{m_Y}}  (\alpha_{-m_T}^T)^{n^T_{m_T}}  
|0;k\rangle
\end{equation}
and satisfies
$N_{\mu, m_\mu}|\vec{\bm{n}};k\rangle= n^\mu_{m_\mu} m_\mu |\vec{\bm{n}};k\rangle $
and $ p|\vec{\bm{n}};k\rangle=k|\vec{\bm{n}};k\rangle.$

\vspace{5mm}
With these notations, we can evaluate  traces of the form: 
\begin{equation}
\mathrm{Tr}\bigl[e^{-2\pi u H_0}O\bigr]\ =\ {\cal T}e^{-\frac{2r^2}{\pi\alpha'}u}\int \frac{dk}{2\pi}e^{-2\pi\alpha'uk^2}
{\rm tr} [ e^{-2\pi u {\tilde H}_0} {\cal O}]
\end{equation}
where 
\begin{equation}
{\rm tr} [ e^{-2\pi u {\tilde H}_0} {\cal O}] =
\sum_{\vec{\bm{n}}} e^{-2\pi u N(\vec{\bf n})}\langle\vec{\bm{n}};k|O|\vec{\bm{n}};k\rangle
 \label{explicit-trace}
\end{equation}
and $N(\vec{\bf n}) := \sum_{\mu=T,X,Y}\sum_{m=1}^\infty m\,n^{\mu}_{m}. $
In order to simplify the calculation,
we denote $[O]_D$ as the diagonal element of  $O$ satisfying
\begin{eqnarray}
\langle \vec{\bm{n}};k|O|\vec{\bm{n}};k\rangle 
&=& \langle \vec{\bm{n}};k|[O]_D|\vec{\bm{n}};k\rangle,
\end{eqnarray}
which are sufficient in the calculations  (\ref{explicit-trace}).
The remaining task to calculate the partition function is to calculate the diagonal elements of $H_1$, $H_1^2$ and $H_2$, and then to integrate over the Schwinger parameter $s$.
In Section \ref{DME123}, we obtain $[H_1]_D$, $[H_1^2]_D$ and $[H_2]_D$, which enable us
to calculate ${\rm Tr}[e^{-2\pi s H_0} H_{1,2}]$ and  ${\rm Tr}[e^{-2\pi s H_0} H_{1}^2]$.
Then in Section \ref{main result}, we 
determine the one-loop partition function up to ${\cal O}(v^2)$.

\subsection{Diagonal elements: $[H_1]_D$, $[H_1^2]_D$ and $[H_2]_D$} \label{DME123}
As defined in  (\ref{H_2-def}), 
the operator $H_1$  is the part of the 
operator $V_1$ (\ref{V_1}) that commutes with $H_0$; namely
$H_1|{\rm state} \rangle \neq |{\rm state} \rangle$, but
has the same energy eigenvalues of $H_0$. 
Using the mode expansions given in Appendix \ref{free theory}, it is straightforward to obtain
\begin{equation}
H_1 = -\frac{2i\alpha'}{r}p\sum_{k\ne0}\frac1k\alpha_k^X\alpha_{-k}^Y+\frac{2i}{\pi}\sum_{k\ne0}\frac1k\alpha_{-k}^T\alpha_k^Y. 
   \label{H_1}
\end{equation}
For any $\vec{\bm n}$, the diagonal matrix element
 $\langle\vec{\bm{n}};k|H_1|\vec{\bm{n}};k\rangle$ vanishes and 
 \begin{equation}
 [H_1]_D=0.
 \label{H1D=0}
 \end{equation} 
Thus, we find that the trace $\mathrm{Tr}\bigl[e^{-2\pi \tau H_0}H_1\bigr]$ in (\ref{expand-Tr}) vanishes. 
Actually, this should be the case since the energy spectrum of the rotating open string is independent of the direction of the rotation, and the linear terms in $v$ must vanish. 

To determine $[H_1^2]_D$, we take the square of (\ref{H_1}) and collect terms which commute with $H_0$. 
We find 
\begin{eqnarray}
[H_1^2]_D 
&=& \frac{4(\alpha')^2}{r^2}p^2\left( 2N_{XY}(2)+N_X(1)+N_Y(1) \right) \nonumber \\
& & +\frac{4}{\pi^2}\left( 2N_{TY}(2)+N_T(1)+N_Y(1) \right), 
\label{H12D}
\end{eqnarray}
where we defined 
\begin{equation}
N_\mu(x) := \sum_{n=1}^\infty\frac1{n^x}N_{\mu,n}, \hspace{1cm} N_{\mu\nu}(x) := \sum_{n=1}^\infty\frac1{n^x}N_{\mu,n}N_{\nu,n}
\label{NalphaDef}
\end{equation}
for $\mu,\nu=T,X,Y$. 
Note that the unperturbed states $|\vec{\bm n};k\rangle$ are the eigenstates of these operators. 
By using the general formulae in Appendix \ref{App-trace-NXY}, we can sum
over all the excited states of the open string and obtain the trace
$\mathrm{Tr}\bigl[e^{-2\pi s H_0}H_1^2 \bigr]$.

We now show the result of $[H_{2}]_D$. It is much more complicated
since $V_2$ defined by  (\ref{V_2}) contains quartic terms  in the world sheet variables, and
we need to appropriately regularize the infinite sum appearing in the intermediate states. 
We leave  the calculations in Appendix \ref{detail H_2}, and show the final result here.
$[H_2]_D$ is a sum of $[V_2]_D$ in  (\ref{V2-result}) 
and $-\sum_m [V_{1,-m}V_{1,m}]_D /m $ in (\ref{V1V1-result}). Each of them
contains many terms including various divergences. 
However, many terms in (\ref{V2-result}) and
(\ref{V1V1-result}) are miraculously cancelled with each other, and 
the final result turns out to be quite simple;
\begin{eqnarray}
 [H_2]_D &=& [V_2]_D -\sum_{m\ne0}\frac1m[V_{1,-m}V_{1,m}]_D 
\nonumber \\
&=& \frac{\alpha'}{r^2}\left[ 2N_{XY}(2)+N_X(1)+N_Y(1) \right] 
 -\frac2{\pi^2}(N_T(2)-N_Y(2))
\nonumber \\
& & -\frac2{\pi^2}(N_T(2)+N_Y(2))-\frac13\alpha'p^2-\frac2{\pi^2}\zeta(1) \nonumber \\
& &
+\frac{\alpha'}{r^2} \left( N_X(0)+N_Y(0)+N_T(0){+\frac14 \zeta(0)} \right).
   \label{H_2 final}
\end{eqnarray}
Since every term is written in terms of the operators $N_\mu(x)$ and $N_{\mu \nu}(x)$, 
we can use the general formulae in Appendix \ref{App-trace-NXY} to calculate the trace
$\mathrm{Tr}\bigl[e^{-2\pi s H_0}H_2 \bigr]$.

Several comments are in order. 
First,  the last line is proportional to the non-zero modes of $H_0$ including the zero-point energy
$\zeta(0)/4=-3/24.$
Thus it can be interpreted as 
the wave-function renormalization of $T, X$ and $Y$ fields with the following
renormalization factor, $Z^{-1}=1+\alpha' \omega^2$.
In interacting theories,  a one-particle state $a^\dagger|0\rangle$
is no longer an eigenstate of the Hamiltonian and we need to construct a  state 
to include a $Z$-factor so that a state
$Z^{1/2} \left(a^\dagger|0\rangle + |{\rm multi} \rangle \right)$ is a properly normalized eigenstate
of the interacting Hamiltonian.
Here $ |{\rm multi} \rangle$ is a sum of multi-particle states and $Z$ is 
interpreted as the probability of the eigenstate
to be in the single-particle state $a^\dagger|0\rangle.$
In the perturbation theory, $Z$-factor can be read from the renormalization of the free Hamiltonian, and
in the present situation, it is given by
\begin{equation}
 Z^{-1}= 1+ \alpha' \omega^2 >1.
 \label{wave-function-renormalization}
\end{equation}
The same wave-function renormalization factor appears in a simpler calculation of the one-loop
partition function of D0-branes at a constant relative motion discussed in Sec.\ref{linear}.
Since the wave function renormalization gives ${\cal O}(v^2)$ correction to the coupling parameter $v$, 
the last line of $[H_2]_D$ does not contribute to the calculation of the effective potential up to ${\cal O}(v^2)$.
Thus we should replace Eq. (\ref{H_2 final}) with 
\begin{eqnarray}
 [H_2]_D 
&=& \frac{\alpha'}{r^2}\left[ 2N_{XY}(2)+N_X(1)+N_Y(1) \right] 
-\frac4{\pi^2}N_T(2) \nonumber \\
& & -\frac13\alpha'p^2-\frac2{\pi^2}\zeta(1)
\label{H2Dwfrenormalize}
\end{eqnarray}
in the following calculations.

Second, the final result turned out to be very simple after 
the miraculous cancellations between (\ref{V2-result}) and (\ref{V1V1-result}). 
Especially, the operator-valued terms with divergent coefficients, for example (\ref{divergent}), cancel completely. 
This cancellation might be intimately related to the 
the renormalization property of the non-linear sigma model (\ref{action}). 
At the one-loop level, the beta function of the target space metric is proportional to the Ricci tensor of the metric. 
Since the background metric of (\ref{action}) is flat and Ricci tensor vanishes, 
the background metric of the action should not be renormalized 
even though divergence appears in the intermediate steps of the renormalization procedure. 

Finally, the only remaining divergence appears in the  zero-point energy which is
 independent of $\alpha' /r^2 $ in $H_2$. 
This zero-point energy 
must be also renormalized to obtain a sensible mass spectrum of the rotating open string, and
we need to find the correct renormalization scheme to fix the finite part, say $\epsilon_0$, of $H_2$;
\begin{equation}
  -\frac2{\pi^2}\zeta(1) \longrightarrow  \epsilon_0.
\end{equation} 
One possible way to fix $\epsilon_0$ will be
 to check the BRST algebra of the worldsheet theory (\ref{action}), as it determines the intercept for strings in the Minkowski space-time. 
Another possible way will be to examine the behavior of the one-loop partition function in the closed string channel, or in other words, at large distances. 
There must be the contribution from the massless graviton exchanged between the D0-branes which must give us the Newton potential. 
Since the large distance behavior of the partition function would depend on $\epsilon_0$, 
the requirement for reproducing the Newton potential may choose the correct value for $\epsilon_0$. 
In the following, we leave $\epsilon_0$ to be an unknown parameter. 

\subsection{One-loop open string partition function} \label{main result}
By using
Eq. (\ref{expand-Tr}), Eqs. (\ref{H1D=0}), (\ref{H12D}), and (\ref{H2Dwfrenormalize}), and various 
formulae derived in Appendix \ref{App-trace-NXY}, 
we can calculate the one-loop open string partition function (\ref{F}).
Many terms are miraculously cancelled and we have a very simple form 
\begin{eqnarray}
Z 
&=& {\cal T}\int_0^\infty\frac{ds}{2s}(8\pi^2\alpha's)^{-\frac12}e^{-\frac{2r^2}{\pi\alpha'}s}\eta(is)^{-24}\left( 1-\frac13v^2 \right)^{-\frac12} \nonumber \\
& & \times\left[ 1-2\pi v^2\left( -\frac4{\pi^2}s\sum_{n=1}^\infty\frac{n^{-1}q^n}{1-q^n}+\epsilon_0s-\frac4{\pi}s^2\sum_{n=1}^\infty\frac{2q^n}{(1-q^n)^2} \right) \right]+{\cal O}(v^4), \nonumber \\
   \label{partition function final}
\end{eqnarray}
where $q:=e^{-2\pi s}$ and $\eta(is)=q^{1/24}\prod_{m=1}^\infty (1-q^m)$. 
The derivation of this expression is summarized in Appendix \ref{App-trace-NXY}. 
In the $v \rightarrow 0$ limit,  Eq. (\ref{partition function final}) is reduced to 
the partition function in Eq. (\ref{partition-function-atrest}) for D0-branes at rest.
Then let us compare Eq. (\ref{partition function final}) with the 
partition function for D-branes moving with a constant relative velocity $2v$, which is given 
 in Eq. (\ref{exact 1-loop}). 
Some of the $v^2$-corrections in Eq. (\ref{partition function final}) 
that come from the non-zero modes, namely
the third term of the $v^2$-corrections in the square bracket in Eq. (\ref{partition function final}), 
are exactly the same as the $v^2$-corrections in the constant-velocity case in Eq. (\ref{exact 1-loop}).
This term came from the ${\cal O}(v)$-mixing of $T$ and $Y$ in (\ref{H_1}), which exists
in both systems. 
Eq. (\ref{partition function final}), however,  contains more $v^2$-corrections. 
That is, 
the first and the second terms in the curly bracket in Eq. (\ref{partition function final})
 are peculiar only for the revolving case, and do not exist in Eq. (\ref{exact 1-loop}).
   Thus we can say that 
the partition function for the revolving case contains not only 
the corrections due to the velocity $v$ but also 
corrections due to the acceleration $\omega$. 


\section{Effective potential at short distance} \label{effective potential}

The one-loop partition function (\ref{partition function final}) of the open string
gives the effective potential ${\cal V}(r,\omega)=-Z/{\cal T}$ between the two D0-branes induced by the exchange of a single closed string. 
In the present calculation,  we are interested in the short distance behavior of the potential, 
namely $r \ll \sqrt{\alpha'}$. We  expand the potential  as a sum
\begin{equation}
 {\cal V}(r,\omega)\  = 
\ \sum_{n=-1}^\infty {\cal V}_n(r,\omega), 
\end{equation}
where each term ${\cal V}_n(r,\omega)$ corresponds to the contribution of the open string states with the mass level $n+1$ to the partition function. 
$n$ corresponds to the power of $q$ in the $q$-expansion of the integrand of (\ref{partition function final}). 
Therefore, e.g., ${\cal V}_{-1}(r,\omega)$ is the contribution from the tachyon, and ${\cal V}_0(r,\omega)$ comes from the states which are massless when $v=r=0$, and ${\cal V}_{n\ge1}(r,\omega)$ are those from massive open string
states. In the following, we  ignore the tachyon contribution ${\cal V}_{-1}(r,\omega)$. 

\subsection{Massive contributions: ${\cal V}_{n\ge1}(r,\omega)$}
First let us consider contributions of the massive open string states.
It is rather straightforward to evaluate the massive contributions ${\cal V}_n(r,\omega)$ with 
$n\ge1$. 
For example, by expanding the partition function (\ref{partition function final})
with respect to $q$ and take the linear terms in $q$, we obtain ${\cal V}_1(r,\omega)$ as
\begin{eqnarray}
{\cal V}_1(r,\omega) 
&=& -\int_0^\infty\frac{ds}{2s}f(s,r)e^{-2\pi s} \left[ 324+\left( \frac{204}{\pi}-648\epsilon_0 \right)v^2s+432v^2s^2 \right] \nonumber \\ [2mm]
& & +{\cal O}(v^4), 
\end{eqnarray}
where 
\begin{equation}
 f(s,r) := \left( 1-\frac{1}{3}v^2 \right)^{-\frac12}\left( 8\pi^2\alpha's \right)^{-\frac12}e^{-\frac{2r^2}{\pi\alpha'}s} .
\end{equation}
The term $e^{- \frac{2r^2}{\pi\alpha'}s  }$ is nothing but the effect of stretched 
open strings with distance $2r$, and the additional factor $\left( 1-\frac{1}{3}v^2 \right)^{-\frac12}$
comes from the $v^2$ correction to the momentum integration.

To determine ${\cal V}_1(r,\omega)$, we use the following formulae 
\begin{eqnarray}
& & -\int_0^\infty \frac{ds}{2s}\,f(s,r)e^{-2\pi s}s^k \nonumber \\
&=&  -\left( 1-\frac{1}{3}v^2 \right)^{-\frac12}\left( 16\pi\alpha' \right)^{-\frac12}(2\pi)^{-k}\Gamma\left( k-\frac12 \right)\left( 1+\frac{r^2}{\pi^2\alpha'} \right)^{-k+\frac12}. \nonumber \\
   \label{integral}
\end{eqnarray}
The integral with $k\le\frac12$ is defined by the analytic continuation for $k$. 
Using this formulae, we obtain 
{
\begin{eqnarray}
{\cal V}_1(r,\omega) 
&=&\frac{162}{\sqrt{\alpha'}}\left[ 1+\frac12\left( 1+\frac{-13+9\pi^2+27\pi\epsilon_0}{27}\alpha'\omega^2 \right)\left( \frac{r^2}{\pi^2\alpha'} \right) \right. \nonumber \\
& & \left. \hspace{1cm} -\frac18\left( 1+\frac{-44-18\pi^2+57\pi\epsilon_0}{27}\alpha'\omega^2 \right)\left( \frac{r^2}{\pi^2\alpha'} \right)^2 \right] \nonumber \\
& & +{\cal O}(\omega^4,r^6). 
\end{eqnarray}
}
The potential ${\cal V}_1(r, \omega)$ is a generalization of the potential at rest in Eq. (\ref{V2atrest}).
Since the energy eigenvalue of the first excited massive states are split by the interactions, 
${\cal V}_1(r,\omega)$ is a sum of various contributions from the hypersplitted states. 
Note also 
that since our calculation is performed in the Wick rotated metric, the potential in the Lorentzian metric is obtained by replacing $\omega^2$ with $-\omega^2$.
The potential correctly reproduces the static limit (\ref{V2atrest}) at $\omega=0$.
The relevant part of the potential  is written as a positive power series of
$(r/l_{\text str})^2$ and $(\omega/m_{\text str})^2$;
\begin{eqnarray}
{\cal V}_2 &\sim& 
 m_{\text str}  \left( c_1 + c_2 \left( \frac{w}{m_{\text str}} \right)^2 + \cdots  \right) 
\left( \frac{r}{ l_{\text str}} \right)^2  \nonumber \\ 
&+&   m_{\text str} \left(c_3 + c_4 \left( \frac{w}{m_{\text str}} \right)^2 + \cdots  \right) 
 \left( \frac{r}{ l_{\text str}} \right)^4  +\cdots 
 \label{typicalEP-massive}
\end{eqnarray}
where we defined $(2\pi \alpha')^{1/2} = l_{\text str} =1/m_{\text str}$ and $c_1>0$. 
{As expected, the leading order potential is proportional to $m_{\text str}^3 r^2$
and  strongly attractive \cite{Kitazawa}. }
The second and higher terms in each bracket give angular-frequency corrections. 
The leading $\omega^2$ correction to the effective potential 
is given by $m_{\text str} \omega^2 r^2.$  
In the superstring case, the potential must vanish in the $\omega \rightarrow 0$ limit 
where D-branes are at rest.
Thus this $\omega$-dependent term gives the leading order massive state contribution to the potential. 

Other contributions ${\cal V}_n(r,\omega)$ with $n\ge2$ can be obtained similarly.

\subsection{Massless contributions; ${\cal V}_0(r,\omega)$}   \label{sec-diagonalization}

Next, let us consider contributions from massless open string states 
${\cal V}_0(r,\omega)$ given by the $q^0$ terms in the expansion of 
the partition function (\ref{partition function final}); 
\begin{equation}
{\cal V}_0(r,\omega)\ =\ -\int_0^\infty\frac{ds}{2s}\,f(s,r)\left[ 24+\left( \frac8\pi-48\pi\epsilon_0 \right)v^2s+16v^2s^2 \right]+{\cal O}(v^4). 
   \label{V_1(r,omega)}
\end{equation}
It is a sum of contributions from massless vector bosons $\alpha_{-1}^\mu|k\rangle$.
This type of contributions to the effective potential, in particular in the case
of D3-branes,  would correspond to  the 
Coleman-Weinberg type effective potential 
since its mass is given by the distance (i.e. moduli) between the branes.
 In the current setup, since the mass also depends on the angular frequency, 
the corresponding Coleman-Weinberg potential must be evaluated 
in presence of time-dependent scalar expectation value. 
Note that the result of the integral in (\ref{V_1(r,omega)}) is singular at $r=0$ and
the determination of ${\cal V}_0(r,\omega)$ needs some care. 

The effective potential ${\cal V}_0(r,\omega)$ can be obtained if one knows
the  first four eigenvalues of $H_0(v)$, which we denote them by $E_i(k,r,\omega)$ ($i=0,\cdots,3$);
\begin{equation}
{\cal V}_0(r,\omega)\ = -
\ \int_0^\infty\frac{ds}{2s}\int\frac{dk}{2\pi}\left[ \sum_{i=1}^3e^{-2\pi s(E_i(k,r,\omega)-1)}+21e^{-2\pi sE_0(k,r,\omega)} \right]. 
\label{V1tk-integration}
\end{equation}
Since $H_0(v)$ commutes with $H_0$, these eigenvalues can be obtained by diagonalizing the upper-left four-by-four submatrix for $H_0(v)$. 
{ They are given in Appendix \ref{sec-matrixelement} and summarized as }
\begin{eqnarray}
H_0(v) 
&=& \left( 1-\frac13v^2 \right)\alpha'k^2+\frac{r^2}{\pi^2\alpha'}+1+v^2\epsilon_0 \nonumber \\ [2mm]
& & +\left[ 
\begin{array}{cccc}
\displaystyle{-1} & 0 & 0 & 0 \\
0 & \displaystyle{-\frac4{\pi^2} v^2} & \displaystyle{-\frac{2\alpha'}{\pi r}kv^2} & \displaystyle{\frac{2i}{\pi}v} \\ [4mm]
0 & \displaystyle{-\frac{2\alpha'}{\pi r}kv^2} & \displaystyle{\frac{\alpha'}{r^2}v^2} &  \displaystyle{\frac{2i\alpha'}{r}kv} \\ [4mm]
0 & \displaystyle{-\frac{2i}{\pi}v} & \displaystyle{-\frac{2i\alpha'}{r}kv} & \displaystyle{\frac{\alpha'}{r^2}v^2}
\end{array}
\right]+{\cal O}(v^3), 
\end{eqnarray}
corresponding to the states $|\vec{\bm n};k\rangle$ with $N(\vec{\bm n})\le1$. 

The eigenvalues $E_i(k,r,\omega)$ $(i=0,\cdots,3)$ are given by
\begin{equation}
E_i(k,r,\omega)\ =\ \left( 1-\frac13v^2 \right)\alpha'k^2+\frac{r^2}{\pi^2\alpha'}+1+v^2\epsilon_0 +{\cal E}_i(k,r,\omega), 
\end{equation}
where 
\begin{eqnarray}
{\cal E}_0(k,r,\omega) 
&=& -1+{\cal O}(v^4), \\ [2mm]
{\cal E}_1(k,r,\omega) 
&=& \frac1{\pi^2}v^2h(k,r)^{-1}
+{\cal O}(v^4), \\ [2mm]
{\cal E}_2(k,r,\omega) 
&=& 2v\frac{\sqrt{\alpha'}}{r}h(k,r)^\frac12+\frac1{\pi^2}v^2\left[ \frac{\pi^2\alpha'}{r^2}-2
-\frac12h(k,r)^{-1} \right]
+{\cal O}(v^4),\nonumber \\ \\
{\cal E}_3(k,r,\omega) 
&=& -2v\frac{\sqrt{\alpha'}}{r}h(k,r)^\frac12+\frac1{\pi^2}v^2\left[ \frac{\pi^2\alpha'}{r^2}-2
-\frac12h(k,r)^{-1} \right]
+{\cal O}(v^4), \nonumber \\ 
\end{eqnarray}
and 
\begin{equation}
h(k,r)\ :=\ \alpha'k^2+\frac{r^2}{\pi^2\alpha'}. 
\end{equation}
The integral (\ref{V1tk-integration}) is nothing but the Schwinger parameter representation
of the partition function of $(1+0)$-dimensional particles whose energy is given by $2 \pi E_i$. 

\vspace{5mm}
Now we perform the integrations (\ref{V1tk-integration}).
For the eigenvalue $E_0(k,r,\omega)$, the integral can be performed easily. 
We obtain 
\begin{eqnarray}
{\cal V}_{00} =- \int_0^\infty\frac{ds}{2s}\int\frac{dk}{2\pi}\,e^{-2\pi uE_0(k,r,\omega)}
&=& 
\sqrt{\frac{\displaystyle{ 1+\epsilon_0\pi^2\alpha'\omega^2}}{\displaystyle{4\alpha'\left( 1-\frac13\omega^2 r^2 \right)}}\frac{r^2}{\pi^2\alpha'}}+{\cal O}(v^4). 
\nonumber \\ [2mm] \label{EP-massless1}
\end{eqnarray}
The remaining integrals have more complicated forms. 
For the purpose of a rough estimate, we make the following approximation 
\begin{equation}
h(k,r)^\frac12\to\sqrt{\alpha'}|k|, \hspace{1cm} v^2h(k,r)^{-1}\to0. 
\end{equation}
In this approximation, we obtain 
\begin{eqnarray}
{\cal V}_{01}= - \int_0^\infty\frac{ds}{2s}\int\frac{dk}{2\pi}\,e^{-2\pi s(E_1(k,r,\omega)-1)}
&\to&
\sqrt{\frac{\displaystyle{1+\epsilon_0\pi^2\alpha'\omega^2}}{\displaystyle{4\alpha'\left( 1-\frac13\omega^2 r^2 \right)}}\frac{r^2}{\pi^2\alpha'}}, \nonumber \\ 
\label{EP-massless2}
\end{eqnarray}
\begin{eqnarray}
{\cal V}_{02} +{\cal V}_{03}
&= & - \int_0^\infty\frac{ds}{2s}\int\frac{dk}{2\pi}\,e^{-2\pi s(E_2(k,r,\omega)-1)}
- \int_0^\infty\frac{ds}{2s}\int\frac{dk}{2\pi}\,e^{-2\pi s(E_3(k,r,\omega)-1)} \nonumber \\ [2mm]
&\to&
\sqrt{
\frac{\displaystyle{ 1+\left( \epsilon_0\pi^2-2 \right)\alpha'\omega^2 }}{\alpha' \displaystyle{\left( 1-\frac13\omega^2 r^2 \right)}}\displaystyle{\frac{r^2}{\pi^2 \alpha'}}
}. 
\label{EP-massless3}
\end{eqnarray}
The details of the integrations are given in Appendix \ref{massless V}.

The effective potential induced by the massless modes becomes a sum of 
(\ref{EP-massless1}), (\ref{EP-massless2}) and (\ref{EP-massless3}), 
\begin{equation}
{\cal V}_{0} = \sum_{i=0}^{3} {\cal V}_{0i}
\end{equation}
and
in the Lorentzian metric,  $\omega^2$ is replaced with $-\omega^2$. 
In the limit, $r \ll l_{\text str}$ and $\omega \ll m_{\text str}$, 
each term is written in the  form of
\begin{equation}
 m_{\text str} \sqrt{ \left(\frac{r}{l_{\text str}} \right)^2 
 + C \left(\frac{r}{l_{\text str}} \right)^2 \left( \frac{\omega^2}{m_{\text str}} \right)^2 
}
\label{EP-masslesstypical}
\end{equation}
At $\omega=0$, the potential becomes $r/l_s^2$, which is proportional to the 
length of the stretched string.

The typical form of the effective potential induced by the massless state
 (\ref{EP-masslesstypical}) is nothing but the Coleman-Weinberg potential for quantum particles
 in $(1+0)$-dimensions, which can be seen as follows.
 For comparison and future generalizations to D$p$-branes,
 we will consider general cases in $(1+p)$-dimensions. 
In $(1+p)$-dimensional field theory, one-loop integral of a scalar field with mass $m$ is given by
\begin{eqnarray}
- {\text Tr} \log (\Delta+m^2)^{-1/2} 
&=&  \int_\epsilon^\infty \frac{ds}{2s} \int \frac{d^{p+1}k}{(2\pi)^{p+1}} e^{-(k^2+m^2)s} \nonumber \\
&\sim&  \int_\epsilon^\infty \frac{ds}{2s} s^{-\frac{p+1}{2}} e^{-m^2 s} 
\sim
 \left\{ 
\begin{array}{ll}
(m^2)^{\frac{p+1}{2}} & \  p={\text even } \\
(m^2)^{\frac{p+1}{2}} \log m^2 &  \  p={\text odd }.
\end{array}   \right.
\nonumber \\
\label{CW-relation}
\end{eqnarray} 
If mass is given by  vacuum expectation value
 of some scalar field $\phi$, it gives the well-known Coleman-Weinberg potential.
In our case, mass is generated by the distance $r$ between D0-particles and the angular velocity $\omega.$
Thus  the effective potential induced by open string
massless states is given by the typical form  (\ref{EP-masslesstypical}), namely 
$p=0$ case in (\ref{CW-relation}).

\section{D0-branes at constant velocities} \label{linear}
In this section, we study a system of D0-branes at a constant relative velocity for comparison with
the revolving case. 
One might think that the effective potential we obtained in the last section could be derived in a simpler manner
by considering a D-brane system with a constant relative velocity. 
Namely, one may consider a system of two D0-branes whose trajectories are given by
\begin{equation}
\left\{
\begin{array}{l}
x_1(t)\ =\ r, \\ [2mm]
y_1(t)\ =\ vt, 
\end{array}
\right. \hspace{1cm} \left\{
\begin{array}{l}
x_2(t)\ =\ -r, \\ [2mm]
y_2(t)\ =\ -vt.  
\end{array}
\right.
   \label{linear trajectory}
\end{equation}
In the following, we refer this D0-brane system as the linear system, and the revolving D0-branes discussed so far as the revolving system. 
At the moment $t=0$, the kinematic configuration of the linear system is the same as that of the revolving system, as far as the distance and the velocity are concerned. 
Therefore, one might expect that the effective potential for the revolving system would be obtained from the linear system at $t=0$, at least at the order of perturbation we performed in the previous sections, and
the corrections to the effective potential might coincide between these two systems. 
This is not the case 
since the effect of one D0-brane propagates with a finite speed and
the effective potential depends on details of the trajectories of the other D0-brane at $t<0$. 
Thus the interaction will depend not only on the velocity but on the acceleration at the moment of $t=0.$
This was shown in the closed string channel in \cite{Kazama:1997bc}.
In this section, we will show  similarities and  differences between these two systems in the open string channel.

\vspace{5mm}

Although the theory can be solved exactly in terms of a twisted boson\cite{Polchinski:1998rr},
it is instructive to solve the worldsheet theory for the linear system in a non-trivial coordinate system
similar to the rotational coordinate.
We will observe that the resulting action has a  similar form to the action (\ref{action}) for the revolving system,
which helps us to compare the two systems. 

Consider the worldsheet theory of an open string with a boundary condition given by (\ref{linear trajectory}). 
We introduce the following coordinate system (in the Euclidean signature): 
\begin{eqnarray}
x' &=& x, \\
y' &=& y\cos \omega x-t\sin\omega x, \\
t' &=& y\sin\omega x+t\cos\omega x.  
\end{eqnarray}
In this new coordinate system, the trajectories (\ref{linear trajectory}) of the D0-branes become simply 
\begin{equation}
\left\{
\begin{array}{l}
x'_1(t)\ =\ r, \\ [2mm]
y'_1(t)\ =\ 0, 
\end{array}
\right. \hspace{1cm} \left\{
\begin{array}{l}
x'_2(t)\ =\ -r, \\ [2mm]
y'_2(t)\ =\ 0, 
\end{array}
\right.
   \label{trajectory_constant}
\end{equation}
provided that $\omega$ satisfies 
\begin{equation}
v\ =\ \tan r\omega. 
\end{equation}
Note that this relation between $v$ and $\omega$ is the same as the one (\ref{orbit})
for the revolving system up to ${\cal O}(v^2)$. 

The worldsheet action in this coordinate system is 
\begin{eqnarray}
S 
&=& -\frac{r^2}{4\pi\alpha'}\int d^2\sigma\left[ \partial_\alpha \tilde{T}\partial^\alpha \tilde{T}+\partial_\alpha \tilde{X}\partial^\alpha \tilde{X}+\partial_\alpha \tilde{Y}\partial^\alpha \tilde{Y}+\partial_\alpha \tilde{X}^i\partial^\alpha \tilde{X}_i \right. \nonumber \\ [1mm]
& & \left. \hspace{1cm}+2v\partial_\alpha \tilde{X}(\tilde{T}\partial^\alpha \tilde{Y}-\tilde{Y}\partial^\alpha \tilde{T})+v^2(\tilde{T}^2+\tilde{Y}^2)\partial_\alpha \tilde{X}\partial^\alpha \tilde{X}+{\cal O}(v^3) \right]. \nonumber \\
\end{eqnarray}
Note that we have rescaled the fields by $r$. 
This is quite similar to the worldsheet action (\ref{action}) for the revolving system. 

The boundary conditions for $\tilde{X}$ and $\tilde{Y}$ are determined by (\ref{trajectory_constant}). 
Define $X$ by 
\begin{equation}
\tilde{X}\ =\ x(\sigma)+X. 
\end{equation}
Then, $X$ and $\tilde{Y}$ obey 
\begin{equation}
X\Big|_{\sigma=0,\pi}\ =\ \tilde{Y}\Big|_{\sigma=0,\pi}\ =\ 0. 
\end{equation}
One can show that $\tilde{T}$ obeys the Neumann boundary condition;
\begin{equation}
 \partial_\sigma \tilde{T}\Big|_{\sigma=0, \pi}=0 .
\end{equation} 
Note that, unlike the revolving system, 
we do not need any field redefinition like (\ref{new-field-T})
to simplify the boundary condition for $\tilde{T}$. 
We define $T$ by 
\begin{equation}
\tilde{T}\ =\ \frac tr+T, 
\end{equation}
where $t$ is the coordinate zero mode of $\tilde{T}$. 
In the following, we use $Y$ instead of $\tilde{Y}$ for notational simplicity. 

The Hamiltonian is given as 
\begin{equation}
H_{\rm linear}\ =\ H^{(l)}_0+vV^{(l)}_1+v^2V^{(l)}_2+{\cal O}(v^4), 
\end{equation}
where 
\begin{eqnarray}
H^{(l)}_0 
&=& \int^\pi_0d\sigma \Biggl[\frac{\pi \alpha'}{r^2}({\Pi}_T^2+{\Pi}_X^2+{\Pi}_Y^2)+\frac{r^2}{4\pi \alpha'}\left\{ (\partial_\sigma{T})^2+(\partial_\sigma{X})^2+(\partial_\sigma{Y})^2 \right\}\Biggr] \nonumber \\
& &+\frac{r^2}{\pi^2\alpha'}+\frac{v^2}{\pi^2\alpha'}t^2, 
\label{H0-constvelocity}
\\
V^{(l)}_1 
&=& \int^{\pi}_{0}d\sigma \Biggl[-\frac{2\pi \alpha'}{r^2}{\Pi}_X( \tilde{T}{\Pi}_Y-{\Pi}_T Y)+\frac{r^2}{2\pi \alpha'}\partial_\sigma  X( \tilde{T}\partial_\sigma Y-\partial_\sigma T Y)+\frac{2r^2}{\pi^2\alpha'}\partial_\sigma TY\Biggr], \nonumber \\ \\
V^{(l)}_2 
&=& \int^{\pi}_{0}d\sigma \Biggl[\frac{\pi \alpha'}{r^2}(\tilde{T}\Pi_Y-Y\Pi_T)^2+\frac{r^2}{4\pi\alpha'}\Biggl\{ (\partial_\sigma X)^2(\tilde{T}^2+Y^2)+\frac4{\pi^2}(T^2+Y^2) \nonumber \\
& & \hspace*{1cm}+\frac{8}{\pi^2r}tT-\frac4\pi\partial_\sigma X(\tilde{T}^2+Y^2)\Biggr\}\Biggr]. 
   \label{V^{(l)}_2}
\end{eqnarray}
Note that we have moved a term proportional to $v^2t^2$ from the interaction part
$V_2^{(l)}$ to the free part $H^{(l)}_0$
since the sum $r^2+v^2 t^2$ is the distance squared between the D0-branes.\footnote {Since the system of
D0-branes at a constant relative motion does not have invariance under time translation, 
the zero-mode of $T$ is no longer decoupled. It is a big difference from the revolving system noted in 
footnote \ref{footnote-0mode}, and we need to compare two systems at $t=0$.}
This is necessary to improve the behavior of the perturbative result for small $v$. 
From this Hamiltonian, using the improved perturbation theory in Appendix \ref{IOS}, we can construct
\begin{equation}
H^{(l)}_0(v)\ :=\ H^{(l)}_0+vH^{(l)}_1+v^2H^{(l)}_2+{\cal O}(v^4) 
\end{equation}
 which is a counterpart of $H_0(v)$ (\ref{H_0(v)}) in the linear system. 
Note that the terms in the second line in (\ref{V^{(l)}_2}) do not contribute to $H^{(l)}_2$,
and we can drop them in the following discussions.

Recall that $V_1$ and $V_2$ in the Euclideanized theory for the revolving system are 
\begin{eqnarray}
V_1&=&\int^{\pi}_{0}d\sigma \Biggl[-\frac{2\pi \alpha'}{r^2}{\Pi}_T( X{\Pi}_Y-{\Pi}_X Y)+\frac{r^2}{2\pi \alpha'}\partial_\sigma  T( X\partial_\sigma Y-\partial_\sigma X Y)+\frac{2r^2}{\pi^2\alpha'}\partial_\sigma TY\Biggr], \nonumber \\ \\
V_2&=&\int^{\pi}_{0}d\sigma \Biggl[\frac{\pi \alpha'}{r^2}\left\{ (X\Pi_Y-Y\Pi_X)^2+x(\sigma)^2\Bigl({\Pi}_Y^2-{\Pi}_T^2\Bigr) \right. \nonumber \\
&&\hspace{1cm}+\frac{r^2}{4\pi \alpha'}\Biggl \{(\partial_\sigma T)^2(X^2+Y^2)+\frac{8}{\pi^2}Y^2+x(\sigma)^2\Bigl((\partial_\sigma T)^2-(\partial_\sigma Y)^2\Bigr)
\Biggr\}\Biggr], \nonumber \\
\end{eqnarray}
where only terms which contributes to $H_1$ and $H_2$ are shown above. 
The interaction terms $V^{(l)}_1$ and $V^{(l)}_2$ have a quite similar structure to their counterparts in the  revolving system. 
However, there are some differences. 

Apart from the potential term $v^2t^2/\pi^2\alpha'$ in the linear system in (\ref{H0-constvelocity}), 
there are two differences between the revolving and the linear systems. 
First, 
 $T$ and $X$ are exchanged in many terms in the interactions $V_1$ and $V_2$.  
Since $T$ and $X$ obey different, Neumann and Dirichlet,  boundary conditions, 
these terms give  different contributions to $[H_2]_D$; namely
the first line of   (\ref{H_2 final})  in the revolving system. 
Second,  $V_2$ in the revolving case has the following terms 
\begin{equation}
\int_0^\pi d\sigma\Biggl[ \frac{\pi\alpha'}{r^2}x(\sigma)^2(\Pi_Y^2-\Pi_T^2)+\frac{r^2}{4\pi\alpha'}x(\sigma)^2\bigl( (\partial_\sigma T)^2-(\partial_\sigma Y)^2 \bigr) \Biggr], 
   \label{difference in V_2}
\end{equation}
which do not exist in the linear case $V^{(l)}_2$. 
The diagonal part of (\ref{difference in V_2}) which contributes to $H_2$ is given by the 
second line of (\ref{H_2 final}) in $[H_2]_D$ in the revolving system. 
Therefore, the results for the linear system are actually different from those for the revolving system. 
There are other terms in $V_2$ which are absent in $V^{(l)}_2$, but they do not contribute to the results up to ${\cal O}(v^2)$. 

After straightforward calculations in Appendix \ref{linear H_2}, 
we obtain 
\begin{eqnarray}
H^{(l)}_0 
&=& \alpha'p^2+\frac{v^2}{\pi^2\alpha'}t^2+\frac{r^2}{\pi^2\alpha'}+\sum_{n=1}^\infty(N_{T,n}+N_{X,n}+N_{Y,n})-\frac18, 
\\
H^{(l)}_1 
&=& \frac{2i}{\pi}\sum_{n\ne0}\frac1n\alpha^T_n\alpha^Y_{-n}.
\label{Hlinear1}
\end{eqnarray}
For $[H_2^{(l)}]_D$  (see Appendix \ref{linear H_2}), 
most of the terms are cancelled and  the following terms for the wave function renormalization
\begin{equation}
 [H^{(l)}_2]\ =\ \frac{\alpha'}{r^2}  \left( \sum_n N_{T,n} + N_{X,n} +N_{Y,n} -\frac{3}{24}\right)
\end{equation}  
remains. It is the same as  (\ref{wave-function-renormalization}). 
Once the wave function is renormalized, 
it does not affect the final results of the partition function up to  ${\cal O}(v^2)$. 
We have seen that the perturbative expansion of the linear system is
 similar to the revolving systems, but the above differences lead to 
the apparently different results. 

Let us calculate the partition function of the linear system using $H^{(l)}_0(v)$. 
As we mentioned above, $H^{(l)}_0$ includes the term proportional to $t^2$. 
Therefore, the eigenvalues of the zero modes, $t$ and $p$,  no longer take continuous values. 
Instead, they form a harmonic oscillator with the angular frequency $2v/\pi$. 
The effect of $H^{(l)}_1$ in Eq. (\ref{Hlinear1})
to the partition function can be easily determined by diagonalizing it. 
We find that this gives shifts $\pm 2v/\pi$ to the eigenvalues of $H^{(l)}_0$. 
Since $H^{(l)}_2$ does not contribute to the partition function up to ${\cal O}(v^2)$, we see
 that $H^{(l)}_0(v)$ reproduces the correct partition function of the twisted boson\cite{Polchinski:1998rr}
\begin{eqnarray}
Z 
&=& \int_0^\infty\frac{ds}{2s}\,q^{-1}\frac{q^{\frac{v}{\pi} }}{1-q^{\frac{2v}\pi}}e^{-\frac{2r^2}{\pi\alpha'}s}\prod_{n=1}^\infty\Bigl( 1-q^{n+\frac{2v}\pi} \Bigr)^{-1}\Bigl( 1-q^{n-\frac{2v}\pi} \Bigr)^{-1}\Bigl( 1-q^n \Bigr)^{-22} \nonumber \\
&=& \int_0^\infty\frac{ds}{2s}\,     
\frac{-2}{\sinh(2vs)}
e^{-\frac{2r^2}{\pi\alpha'}s}\eta(is)^{-24}\left[ 1-2\pi v^2\left( -\frac4\pi s^2\sum_{n=1}^\infty\frac{2q^n}{(1-q^n)^2} \right) \right]+{\cal O}(v^4) \nonumber \\
   \label{exact 1-loop}
\end{eqnarray}
up to ${\cal O}(v^2)$. 
This can be regarded as a consistency check of the calculation and
 the validity of the formalism we employed in this paper. 
The $v^2$ corrections in the linear system give the third term in the $v^2$ corrections in the 
revolving system in Eq. (\ref{partition function final}). 
Note that the partition function itself is a function of $v^2$, but the energy eigenvalues
are shifted by $\pm 2v/\pi$. This comes from the ${\cal O}(v)$-mixing between 
the world sheet variables, $T$ and $Y$. 
Also note that after Wick rotation to the Lorentzian metric $v \rightarrow iv$, the integrand has
zeros at integer values of $2vs/\pi$, 
which generates an imaginary part in the partition function. 
It reflects open string pair creation \cite{Bachas:1995kx}. 
Such imaginary part does not exist for the revolving system since the distance between D-branes
are constant and non-adiabatic particle creation does not occur.

\section{Conclusions and discussions} \label{discussion}

We have calculated  one-loop partition function of 
 an open string stretched between two D0-branes which revolve around each other. 
To quantize the corresponding worldsheet theory, we introduced new fields which satisfy simple boundary conditions. 
In compensation, it makes the worldsheet action nonlinear. 
We analyze the resulting system perturbatively, using the formalism developed in \cite{Iso:2017hvi}. 
One of the advantages of using the method of the improved perturbation
 is the fact that $H_0(v)$, containing all perturbative information on the energy spectrum, 
 commutes with $H_0$ by construction. 
This enables us to separate the massless contributions from the massive ones, and also to expand the trace as in (\ref{expand-Tr}). 
After lengthy calculations, we obtained the partition function in Eq. (\ref{partition function final}).
The result shows that it is indeed different from the linear system in which D0-branes are moving 
at a constant relative velocity, but the final formulae is surprisingly simple. 
It is one of the main results in the paper. 

We then calculated the effective potential between such a pair of D0-branes 
in the bosonic case
 when their separation  is shorter than the string length. 
Here we summarize the basic structure of the effective potential. 
It has various contributions from massless and massive
open string states and is  given by the following  form
\begin{eqnarray}
{\cal V} &\sim & m_{\text str} \left[ 
\sum_{i=0}^3  
\sqrt{\left( 1+
  C_i \left(\frac{\omega}{m_{\text str}}\right)^2  \right)
 \left(\frac{r}{l_{\text str}}\right)^2  }
+ \left( c_1 + c_2 \left(\frac{\omega}{m_{\text str}}\right)^2 \right) \left(\frac{r}{l_{\text str}}\right)^2
\right] \nonumber \\
&& + {\cal O}(\omega^4, r^4)
\label{resultD0}
\end{eqnarray}
where  $c_1=162 \sqrt{2/\pi} >0.$
The result shows that there is an attractive force between the rotating bosonic D0-branes. 
A contribution from the open string tachyon is neglected in this expression.

One of our motivation for this calculation is to investigate towards a possibility 
of a revolving D$p$-brane system to make a resonant bound state at short distance 
 \cite{Kabat:1996cu}\cite{Danielsson:1996uw}\cite{Iso:2015mva}. 
Suppose that
 a D$p$-brane  with finite volume $V$ and mass $M = g_s^{-1}m_{\text str}^{p+1} V$  
rotates with an angular momentum $ L= M r^2 \omega.$ 
Then the centrifugal repulsive force is given by 
$ M r \omega^2 $
and under the angular momentum conservation, the centrifugal potential becomes
$ V_{\text L} = L^2/2Mr^2$ .
As we saw in the case of D0-particles in the bosonic string, 
strong attractive potential is generated by the one-loop open string amplitude:
$c_1 m_{\text str}^3  r^2$ for $r \ll l_{\text str}$. 
In the D$p$-brane case, an integration over the position of the end point of open strings gives 
the volume factor $V \propto { g_s} M$ and the most dominant part of the effective potential 
will be given by 
$c { g_s} M (r/l_{\text str})^2$ for $r \ll l_{\text str}$. 
Under the angular momentum conservation, $\omega^2$  is replace by $(J^2/Mr^2)^2$
and the centrifugal potential is given by $L^2/2Mr^2$.
When the attractive potential balances with the repulsive centrifugal potential,
 we have a classical solution with the angular velocity $\omega \sim { g_s^{1/2}} m_{\text str}$.
However, such a classical bound state with the large angular velocity $\omega$
is quantum mechanically unstable against emitting closed string emission. Furthermore, 
if we lived on such a D3-brane, the Lorentz symmetry is strongly violated\cite{Iso-Kitazawa};
 thus
a construction of the standard model based on this kind of D-brane configurations in the bosonic string
would not be phenomenologically viable. 
In addition, if $\omega \sim { g_s^{1/2}} m_{\text str}$, we need to take into account 
the effect of background fields which makes the revolving motion on-shell, as mentionted
at the beginning of section \ref{worldsheet}. 
 These corrections are expected to  modify the dimensionsless constants $c, c_1$. 

{In the case of D-branes in the superstring theory, the situation will become different.}
Suppose we start from a BPS configuration such as a pair of D$p$-branes at rest.
When $\omega=0$, such configurations of D-branes are BPS, and therefore, there is no force between them. 
If they are revolving around each other, 
it would generate attractive potential as in (\ref{resultD0}).
But in superstrings, many terms are cancelled  due to the supersymmetry; especially
$\omega$-independent terms are completely cancelled. 
Thus in the typical form of the effective potential generated by massive states (\ref{typicalEP-massive}),
some of the coefficients must vanish: e.g. $c_1=c_3=0.$ 
It is further known that when two branes are moving with a constant relative velocity, 
the $v^2$ terms also cancel\cite{Douglas:1996yp}\cite{Lifschytz:1996iq}.
For revolving D$p$-branes (for odd $p$), massless open string states induce the following
Coleman-Weinberg (CW) type effective potential;
\begin{equation}
V \sum_i  (-1)^{F_i} n_i
 \left( \left(\frac{r}{l_{\text str}}\right)^2 + C_i \left(\frac{\omega}{m_{\text str}}\right)^2 \right)^{\frac{p+1}{2}}
\log \left( \left(\frac{r}{l_{\text str}}\right)^2 + C_i \left(\frac{\omega}{m_{\text str}}\right)^2  \right)
\end{equation}
where $n_i$ is the number of degrees of freedom of the field $i$.
Supersymmetry may impose  $\sum_i (-1)^{F_i} n_i =0$.
{ If the attractive potential is dominated by the CW potential, the balancing condition with 
the centrifugal potential becomes different from the bosonic case, and a possibility of a bound state with 
much lower angular frequency would arise. It is because the attractive potential
in the superstirng case is much weaker than 
the bosonic one, and we might expect }
\begin{equation}
\frac{\omega}{m_{\text str}} , \  \frac{r}{l_{\text str}} \ll 1.
\end{equation} 
If such a bound state can be shown to exist, 
it would  escape the problems of 
strong Lorentz violation as well as the rapid closed string emissions
in phenomenological applications. 
We hope to come back to this issue in future. 

\vspace{1cm}

{\Large \bf Acknowledgements}

\vspace{3mm}

We would like to thank Noriaki Kitazawa for enlightening discussions.
This work  is supported
in part by Grants-in-Aid for Scientific Research (No. 16K05329) and (No. 18H03708) from the
Japan Society for the Promotion of Science. 

\appendix

\section{Improved perturbation theory} \label{IOS}

In this appendix, we briefly review the improved perturbation theory developed in \cite{Iso:2017hvi}. 
Consider a Hamiltonian of the form 
\begin{equation}
H = H_0+\lambda V_1+\lambda^2V_2. 
\end{equation}
In the interaction picture, we regard $H_0$ as the free part and the remaining part as the perturbation. 
Instead, we decompose $H$ as 
\begin{equation}
H = H_0(\lambda)+\lambda V(\lambda), 
\end{equation}
where 
\begin{eqnarray}
H_0(\lambda) 
&:=& H_0+\lambda H_1+\lambda^2H_2+  \lambda^3 H_3 + \cdots  \\ 
V(\lambda) 
&:=& V_1-H_1+\lambda(V_2-H_2) - \lambda^2 H_3 - \cdots 
\end{eqnarray}
The operators $H_1,H_2, H_3$ etc. are to be determined later. 
One can construct a perturbation theory based on this decomposition
up to any desired orders of perturbation with respect to $\lambda$.
The time evolution of operators are given by a unitary operator $U(\lambda,t)$ which satisfy 
\begin{equation}
\frac d{dt}U(\lambda,t) = -i\lambda V(\lambda,t)U(\lambda,t), \hspace{1cm} V(\lambda,t) := e^{iH_0(\lambda)t}V(\lambda)e^{-iH_0(\lambda)t}. 
\end{equation}
The solution satisfying $U(\lambda,0)=I$ is 
\begin{eqnarray}
U(\lambda,t) 
&=& I+(-i\lambda)\int_0^tdt_1\,V(\lambda,t_1) \nonumber \\
& & +(-i\lambda)^2\int_0^tdt_1\int_0^{t_1}dt_2\,V(\lambda,t_1)V(\lambda,t_2) \nonumber \\
& & +(-i\lambda)^3 \int_0^tdt_1\int_0^{t_1}dt_2\int_0^tdt_3 \,V(\lambda,t_1)V(\lambda,t_2)V(\lambda,t_3)
\nonumber \\
& & +\mathcal{O}(\lambda^4). 
\end{eqnarray}
As proved in \cite{Iso:2017hvi}, this operator does not have any secular terms 
provided that $H_n$  are appropriately chosen. 
To show the explicit expressions for them, let us introduce operators $V_{i,a}$ $(i=1,2)$ which satisfy 
\begin{equation}
V_i=\sum_aV_{i,a},\hspace{1cm}[H_0,V_{i,a}]=\omega_aV_{i,a}. 
\end{equation}
Explicitly, they are given as 
\begin{equation}
V_{i,a}=\sum_{E_m-E_n=\omega_a}|m\rangle \langle m|V_i|n\rangle \langle n|, 
\end{equation}
where $|n\rangle$ are the eigenstates of $H_0$. 
In terms of these operators, $H_1$ and $H_2$ are given as 
\begin{eqnarray}
H_1 
&=& V_{1,0}, \\
H_2 
&=& V_{2,0}-\sum_{a,b\ne0}\frac1{\omega_a}\delta_{\omega_a+\omega_b}V_{1,b}V_{1,a}. 
\end{eqnarray}
Note that these operators commute with $H_0$. 
For systematic derivations of higher $H_n$ ($n\ge 3)$, see \cite{Iso:2017hvi}.

As a result, it turns out that the time-dependence of all operators in this system is given in terms of $e^{iH_0(\lambda)t}$ to all orders of perturbation theory. 
Therefore, the eigenvalues of $H_0(\lambda)$ should be the same as those of the full Hamiltonian $H$ to all orders in $\lambda$.

\section{Mode expansions} \label{free theory}

The worldsheet theory (\ref{action}) becomes a free theory when $v=0$. 
In this case, the quantization of the worldsheet fields are straightforward. 
In the Euclidean theory, their mode expansions at $\tau=0$ are as follows: 
\begin{eqnarray}
T 
&=&  \frac tr+i\frac{\sqrt{2\alpha'}}r\sum_{n\ne0}\frac{\alpha^T_n}{n}\cos n\sigma, \\
X 
&=& \frac{\sqrt{2\alpha'}}{r}\sum_{n\ne0}\frac{\alpha^X_n}{n}\sin n\sigma, \\
Y 
&=& \frac{\sqrt{2\alpha'}}{r}\sum_{n\ne0}\frac{\alpha^Y_n}{n}\sin n\sigma, \\
\Pi_T 
&=& \frac{r}{\pi}p+\frac{r}{\pi\sqrt{2\alpha'}}\sum_{n\ne0}\alpha^T_n\cos n\sigma, \\
\Pi_X 
&=& -\frac{ir}{\pi\sqrt{2\alpha'}}\sum_{n\ne0}\alpha^X_n\sin n\sigma, \\
\Pi_Y 
&=& -\frac{ir}{\pi\sqrt{2\alpha'}}\sum_{n\ne0}\alpha^Y_n\sin n\sigma. 
\end{eqnarray}
Note that the radius $r$ appears above because of the rescaling $\tilde X^\mu\to r\tilde X^\mu$ mentioned below (\ref{action-before}). 
The commutation relations of the mode operators are 
\begin{equation}
[ \alpha^T_n,\alpha^T_m ]\ =\ [ \alpha^X_n,\alpha^X_m ]\ =\ [ \alpha^Y_n,\alpha^Y_m ]\ =\ n\delta_{n+m}. 
\end{equation}

\section{Calculation of $[H_2]_D$} \label{detail H_2}
In the appendix, we calculate the diagonal part of $H_2$, which is given as the following sum;
\begin{equation}
[H_2]_D = [V_2]_D -\sum_{m\ne0}\frac1m[V_{1,-m}V_{1,m}]_D .
\end{equation}
The calculations are  complicated
since $V_2$ defined by  (\ref{V_2}) and $(V_1)^2$
contain quartic terms  in the world sheet variables, and
we need to appropriately regularize the infinite sum appearing in the intermediate states.  
Let us first look at  the following term in $V_2$: 
\begin{equation}
\int_0^\sigma d\sigma\,\frac{\pi\alpha'}{r^2}X^2\Pi_Y^2. 
   \label{V_2-sample}
\end{equation}
The diagonal part of this operator is obtained by substituting $X^2$ and $\Pi_Y^2$ with their diagonal parts given by
\begin{eqnarray}
\left[ X(\sigma)^2 \right]_D 
&=& \frac{2\alpha'}{r^2}\sum_{n\ne0}\frac1nD_{X,n}\sin^2 n\sigma, \\
\left[ \Pi_Y(\sigma)^2 \right]_D 
&=& \frac{r^2}{2\pi^2\alpha'}\sum_{n\ne0}nD_{Y,n}\sin^2 n\sigma, 
\end{eqnarray}
where $D_{\mu,n}$ are defined by
\begin{equation}
D_{\mu,n}\ :=\ \frac1nN_{\mu,n}+\theta(n), \hspace{1cm} (\mu=T,X,Y)
\end{equation}
and $\theta(n)$ is the step function;
\begin{equation}
\theta(n) := \left\{
\begin{array}{cc}
1, & (n>0) \\ [2mm]
0. & (n\le0)
\end{array}
\right.
\end{equation} 
A suitable regularization of the summations is necessary.
As a result of the substitution, we obtain (see (\ref{V2D-ex1}))
\begin{eqnarray}
& & \left[\int_0^\pi d\sigma\,\frac{\pi\alpha'}{r^2} X^2 \Pi_Y^2 \right]_D =
\int_0^\pi d\sigma\,\frac{\pi\alpha'}{r^2}[X^2]_D[\Pi_Y^2]_D \nonumber \\
&=& \frac{\alpha'}{4r^2}\sum_{n,m\ne0}\frac mnD_{X,n}D_{Y,m}+\frac{\alpha'}{4r^2}\sum_{n\ne0}D_{X,n}D_{Y,n}
{- \frac{\alpha'}{8r^2}\sum_{n\ne0}D_{X,n} } .
\end{eqnarray}
By using the formulae (\ref{DD-N-formulae} ), we see that the expression contains the following term
\begin{equation}
\frac{\alpha'}{4r^2}
\left(2N_X(2) +\zeta(1)\right)  \left( 2N_Y(0) +  \zeta(-1) \right) , 
   \label{divergent}
\end{equation}
where 
\begin{equation}
\zeta(s) := \sum_{n=1}^\infty\frac1{n^s}
\end{equation}
is the Riemann zeta function. Most of the infinite sums can be regularized  
by using $\zeta(-1)=-1/12$ or $\zeta(0)=-1/2$, but 
$\zeta(1)$ can not be regularized by the zeta function regularization. 
It will be, however, 
observed  that this kind of divergences 
cancels among various terms within $H_2$.

\vspace{5mm}
Other terms in $[V_2]_D$ can be obtained in a similar manner.
Using the diagonal parts of the products given in Appendix \ref{diagonal part},
we find 
\begin{eqnarray}
&& [V_2]_D 
= \frac{\alpha'}{2r^2}\zeta(1)\left( 2N_T(0)+N_X(0)+N_Y(0) \right)
+\frac{\alpha'}{r^2}\zeta(1) \zeta(-1)
\nonumber \\
& & +\frac{\alpha'}{r^2}\left[ N_{XY}(2)+\frac12 (N_{TX}(2)+N_{TY}(2)) \right. \nonumber \\ [2mm]
& & \left. +N_X(2) (N_Y(0) + N_T(0))
+N_Y(2)( N_X(0)+N_T(0) )
\right. \nonumber \\ [2mm]
& & \left. +\zeta(-1)(N_X(2)+N_Y(2))+\frac12N_T(1)+\frac34 (N_X(1)+N_Y(1))
+ \frac12 \zeta(0)
\right]
\nonumber \\
& & -\frac2{\pi^2}(N_T(2)-N_Y(2))-\frac13\alpha'p^2. 
\label{V2-result}
\end{eqnarray}
For details of the calculation, see Appendix \ref{V_2-D}. 

\vspace{5mm}
Next we evaluate the second term in $H_2$  (\ref{H_2-def}).
Since the calculation is similarly performed, it is given in the Appendix \ref{V1V1-D}.
The final result is 
\begin{eqnarray}
&& -\sum_{m\ne0}\frac1m[V_{1,-m}V_{1,m}]_D 
\nonumber \\
&=& -\frac{\alpha'}{2r^2}\zeta(1)\left( 2N_T(0)+N_X(0)+N_Y(0) \right)
-\frac{\alpha'}{r^2}\zeta(1) \zeta(-1)
-\frac2{\pi^2}\zeta(1) \nonumber \\
& & +\frac{\alpha'}{r^2}\left[ N_{XY}(2)-\frac12(N_{TX}(2)+N_{TY}(2)) \right. \nonumber \\ [2mm]
& & \left. -N_X(2)(N_Y(0) + N_T(0))
-N_Y(2)(N_X(0)+N_T(0)) \right. \nonumber \\ [2mm]
& & \left. -\zeta(-1)( N_X(2)+N_Y(2)) -\frac12N_T(1)+\frac14 (N_X(1)+N_Y(1)) \right. \nonumber \\
& & \left. 
+N_X(0)+N_Y(0)+N_T(0)
 +\frac12 \zeta(0)^2 
\right]
-\frac2{\pi^2}(N_T(2)+N_Y(2)). 
\label{V1V1-result}
\end{eqnarray}

\subsection{Diagonal parts} \label{diagonal part}

To derive the explicit form of $[H_2]_D$, we need the diagonal parts of products of fields. 
They are given as follows. 
For products including $T$ and $\Pi_T$, 
\begin{eqnarray}
\left[ \Pi_T(\sigma)\Pi_T(\sigma') \right]_D 
&=& \frac{r^2}{\pi^2}p^2+\frac{r^2}{2\pi^2\alpha'}\sum_{n\ne0}nD_{T,n}\cos n\sigma\cos n\sigma', \\
\left[ \partial_\sigma T(\sigma)\partial_\sigma T(\sigma') \right]_D 
&=& \frac{2\alpha'}{r^2}\sum_{n\ne0}nD_{T,n}\sin n\sigma\sin n\sigma', \\
\left[ \Pi_T(\sigma)\partial_\sigma T(\sigma') \right]_D 
&=& \frac i\pi\sum_{n\ne0} nD_{T,n}\cos n\sigma\sin n\sigma', \\
\left[ \partial_\sigma T(\sigma)\Pi_T(\sigma') \right]_D 
&=& -\frac i\pi\sum_{n\ne0}nD_{T,n}\sin n\sigma \cos n\sigma'. 
\end{eqnarray}
For products including $X$ and $\Pi_X$, 
\begin{eqnarray}
[ X(\sigma)X(\sigma') ]_D 
&=& \frac{2\alpha'}{r^2}\sum_{n\ne0} \frac1nD_{X,n}\sin n\sigma\sin n\sigma', \\
\left[ \Pi_X(\sigma)\Pi_X(\sigma') \right]_D 
&=& \frac{r^2}{2\pi^2\alpha'}\sum_{n\ne0} nD_{X,n}\sin n\sigma\sin n\sigma', \\
\left[ \partial_\sigma X(\sigma)\partial_\sigma X(\sigma') \right]_D
&=& \frac{2\alpha'}{r^2}\sum_{n\ne0}nD_{X,n}\cos n\sigma\cos n\sigma', \\
\left[ X(\sigma)\Pi_X(\sigma') \right]_D 
&=& \frac i\pi\sum_{n\ne0}D_{X,n} \sin n\sigma\sin n\sigma', \\
\left[ \Pi_X(\sigma)X(\sigma') \right]_D 
&=& -\frac i\pi\sum_{n\ne0}D_{X,n} \sin n\sigma\sin n\sigma', \\
\left[ X(\sigma)\partial_\sigma X(\sigma') \right]_D 
&=& \frac{2\alpha'}{r^2}\sum_{n\ne0}D_{X,n}\sin n\sigma\cos n\sigma', \\
\left[ \partial_\sigma X(\sigma)X(\sigma') \right]_D 
&=& \frac{2\alpha'}{r^2}\sum_{n\ne0}D_{X,n} \cos n\sigma\sin n\sigma', \\
\left[ \Pi_X(\sigma)\partial_\sigma X(\sigma') \right]_D 
&=& -\frac i\pi\sum_{n\ne0} nD_{X,n}\sin n\sigma\cos n\sigma', \\
\left[ \partial_\sigma X(\sigma)\Pi_X(\sigma') \right]_D 
&=& \frac i\pi\sum_{n\ne0}nD_{X,n}\cos n\sigma\sin n\sigma'.
\end{eqnarray}
Those including $Y$ and $\Pi_Y$ have the same form as above. 

\subsection{$[V_2]_D$} \label{V_2-D}

The diagonal part of (\ref{V_2-sample}) can be calculated as follows: 
\begin{eqnarray}
& & \int_0^\pi d\sigma\,\frac{\pi\alpha'}{r^2}[X^2]_D[\Pi_Y^2]_D \nonumber \\
&=& \frac{\alpha'}{r^2}\sum_{n\ne0}\frac1nD_{X,n}\sum_{m\ne0}mD_{Y,m}\cdot\frac1\pi\int_0^\pi d\sigma\,\sin^2n\sigma\sin^2m\sigma \nonumber \\
&=& \frac{\alpha'}{4r^2}\sum_{n,m\ne0}\frac mnD_{X,n}D_{Y,m}+\frac{\alpha'}{8r^2}\sum_{n\ne0}\frac1nD_{X,n}\left( nD_{Y,n}+(-n)D_{Y,-n} \right) \nonumber \\
&=& \frac{\alpha'}{4r^2}\sum_{n,m\ne0}\frac mnD_{X,n}D_{Y,m}+\frac{\alpha'}{4r^2}\sum_{n\ne0}D_{X,n}D_{Y,n}-\frac{\alpha'}{8r^2}\sum_{n\ne0}D_{X,n}. 
\label{V2D-ex1}
\end{eqnarray}
We have used the following identities 
\begin{equation}
D_{X,-n}\ =\ -D_{X,n}+1. \hspace{1cm} (n\ne0)
\end{equation}

We introduce the following notations: 
\begin{eqnarray}
D^1_{\mu\nu} 
&:=& \sum_{n,m\ne0}\frac mnD_{\mu,n}D_{\nu,m}, \\
D^2_{\mu\nu} 
&:=& \sum_{n\ne0}D_{\mu,n}D_{\nu,n}, \\
D_\mu(x) 
&:=& \sum_{n\ne0}\frac1{n^x}D_{\mu,n}. 
\end{eqnarray}
In terms of these operators, (\ref{V2D-ex1}) can be written as 
\begin{equation}
\int_0^\pi d\sigma\,\frac{\pi\alpha'}{r^2}[X^2]_D[\Pi_Y^2]_D\ =\ \frac{\alpha'}{4r^2}\left( D^1_{XY}+D^2_{XY}-\frac12D_X(0) \right). 
\end{equation}

The other terms contributing to $[V_2]_D$ are as follows: 
\begin{eqnarray}
\int_0^\pi d\sigma\,\frac{\pi\alpha'}{r^2}[Y^2]_D[\Pi_X^2]_D 
&=& \frac{\alpha'}{4r^2}\left( D^1_{YX}+D^2_{YX}-\frac12D_Y(0) \right), \nonumber \\
-\int_0^\pi d\sigma\,\frac{2\pi\alpha'}{r^2}[X\Pi_X]_D[\Pi_YY]_D 
&=& 0, \nonumber \\
-\int_0^\pi d\sigma\,\frac{2\pi\alpha'}{r^2}[\Pi_XX]_D[Y\Pi_Y]_D 
&=& 0,, \nonumber \\
-\int_0^\pi d\sigma\,\frac{\pi\alpha'}{r^2}x(\sigma)^2[\Pi_T^2]_D 
&=& -\frac{\alpha'}{3}p^2-\frac1{12}D_T(-1)-\frac1{2\pi^2}D_T(1), 
\nonumber 
\end{eqnarray}
\begin{eqnarray}
\int_0^\pi d\sigma\,\frac{\pi\alpha'}{r^2}x(\sigma)^2[\Pi_Y^2]_D 
&=& \frac1{12}D_Y(-1)-\frac1{2\pi^2}D_Y(1), 
\nonumber \\
\int_0^\pi d\sigma\,\frac{r^2}{4\pi\alpha'}[(\partial_\sigma T)^2]_D[X^2]_D 
&=& \frac{\alpha'}{4r^2}\left( D^1_{XT}+D^2_{XT}-\frac12D_T(0) \right),
\nonumber \\
\int_0^\pi d\sigma\,\frac{r^2}{4\pi\alpha'}[(\partial_\sigma T)^2]_D[Y^2]_D 
&=& \frac{\alpha'}{4r^2}\left( D^1_{YT}+D^2_{YT}-\frac12D_T(0) \right),
\nonumber \\
\int_0^\pi d\sigma\,\frac{r^2}{4\pi\alpha'}x(\sigma)^2[(\partial_\sigma T)^2]_D 
&=& \frac1{12}D_T(-1)-\frac1{2\pi^2}D_T(1), 
\nonumber \\
-\int_0^\pi d\sigma\,\frac{r^2}{4\pi\alpha'}x(\sigma)^2[(\partial_\sigma Y)^2]_D 
&=& -\frac1{12}D_Y(-1)-\frac1{2\pi^2}D_Y(1),
\nonumber \\
-\int_0^\pi d\sigma\,\frac{r^2}{4\pi\alpha'}\frac4\pi x(\sigma)[Y\partial_\sigma Y]_D 
&=& -\frac1{\pi^2}D_Y(1),
\nonumber \\
\int_0^\pi d\sigma\,\frac{r^2}{4\pi\alpha'}\frac{12}{\pi^2}[Y^2]_D 
&=& \frac3{\pi^2}D_Y(1). 
\nonumber 
\end{eqnarray}

Summing all of them, $[V_2]_D$ is given as 
\begin{eqnarray}
[V_2]_D 
&=& \frac{\alpha'}{4r^2}\left( D^1_{XY}+D^1_{YX}+D^1_{XT}+D^1_{YT}+2D^2_{XY}+D^2_{XT}+D^2_{YT} \right) \nonumber \\
& & -\frac1{\pi^2}\left( D_T(1)-D_Y(1) \right)-\frac13\alpha'p^2+\frac{\alpha'}{4r^2}. 
\end{eqnarray}
To extract the divergent terms, we use the formulae: 
\begin{eqnarray}
D^1_{XY} 
&=& \zeta(1)\left( 2N_Y(0)  + \zeta(-1) 
 \right)+4N_X(2)N_Y(0)-\frac16N_X(2), \nonumber \\
\label{DD-N-formulae} \\
D^2_{XY} 
&=& 2N_{XY}(2)+N_X(1)+N_Y(1)   + \zeta(0) \\ 
D_T(1)-D_Y(1) 
&=& 2N_T(2)-2N_Y(2), 
\end{eqnarray}
and so on. 
Here we use the zeta-function regularized values: $\zeta(-1)=-1/12$ and $\zeta(0)=-1/2$. 
By using these results, we obtain (\ref{V2-result}). 

\subsection{$[V_1V_1]_D$} \label{V1V1-D}
The diagonal part of the second term  can be calculated as follows. 
For example, let us consider the following term: 
\begin{equation}
-\sum_{m\ne0}\frac1m\int_0^\pi d\sigma\frac{-2\pi\alpha'}{r^2}\left[ \Pi_TX\Pi_Y(\sigma) \right]_{-m}\int_0^\pi d\sigma'\frac{-2\pi\alpha'}{r^2}\left[ \Pi_TX\Pi_Y(\sigma') \right]_{m}
   \label{V_1^2-sample}
\end{equation}
which contributes to $H_2$. 
The diagonal part of this operator can be written as follows: 
\begin{eqnarray}
& & -\frac{2\alpha'}{r^2}\sum_{m,n,k,l\ne0}\frac{nl}{mk}D_{T,n}D_{X,k}D_{Y,l}\delta_{n+k+l,m}\left[ \frac1\pi\int_0^\pi d\sigma\,\cos n\sigma\sin k\sigma\sin l\sigma \right]^2 \nonumber \\
& & -\frac{4(\alpha')^2}{r^2}p^2\sum_{m,k,l\ne0}\frac{l}{mk}D_{X,k}D_{Y,l}\delta_{k+l,m}\left[ \frac1\pi\int_0^\pi d\sigma\,\sin k\sigma\sin l\sigma \right]^2. 
\end{eqnarray}
This expression can be obtained from (\ref{V_1^2-sample}) by substituting $\Pi_T(\sigma)\Pi_T(\sigma')$, $X(\sigma)X(\sigma')$ and $\Pi_Y(\sigma)\Pi_Y(\sigma')$ with their diagonal parts given in Appendix \ref{diagonal part}, and then inserting $\delta_{n+k+l,m}$ and $\delta_{k+l,m}$ at appropriate places in the sums. 
Performing the integrals, we obtain 
\begin{eqnarray}
& & \frac{\alpha'}{8r^2}\sum_{m,n,k,l\ne0}\frac{nl}{mk}D_{T,n}D_{X,k}D_{Y,l}(-\delta_{2l,m}\delta_{l,k+n}-\delta_{2k,m}\delta_{k,l+n}-\delta_{2n,m}\delta_{n,k+l}) \nonumber \\
& & -\frac{(\alpha')^2}{2r^2}p^2\sum_{k\ne0}\frac1kD_{X,k}D_{Y,k}. 
\end{eqnarray}
Again, this contains terms with divergent coefficients. 
For example, 
\begin{eqnarray}
& & -\frac{\alpha'}{8r^2}\sum_{m,n,k,l\ne0}\frac{nl}{mk}D_{T,n}D_{X,k}D_{Y,l}\delta_{2l,m}\delta_{l,k+n} \nonumber \\
&=& -\frac{\alpha'}{16r^2}\zeta(1)\left( 2N_Y(0)  + \zeta(-1)
\right)+\mbox{(finite)}. 
\end{eqnarray}

The other terms can be calculated similarly. 
In the following, we use an abbreviated notation, for example, 
\begin{eqnarray}
&&\left( \Pi_TX\Pi_Y,\Pi_TX\Pi_Y \right) \nonumber \\[2mm]
&:=& -\sum_{m\ne0}\frac1m\int_0^\pi d\sigma\,\frac{-2\pi\alpha'}{r^2}\left[ \Pi_TX\Pi_Y(\sigma) \right]_{-m} \int_0^\pi d\sigma'\,\frac{-2\pi\alpha'}{r^2}\left[ \Pi_TX\Pi_Y(\sigma') \right]_m. \nonumber
\end{eqnarray}
We also use the notations
\begin{eqnarray}
D^{(\pm,\pm,\pm)}_{\mu\nu\rho}
&:=& \sum_{m,n,k,l\ne0}\frac{nl}{mk}D_{\mu,n}D_{\nu,k}D_{\rho,l}\Delta_{(\pm,\pm,\pm)}, \nonumber \\
\tilde{D}^{(\pm,\pm,\pm)}_{\mu\nu\rho} 
&:=& \sum_{m,n,k,l\ne0}\frac{n}{m}D_{\mu,n}D_{\nu,k}D_{\rho,l}\Delta_{(\pm,\pm,\pm)}, \nonumber \\
D^3_{\mu\nu} 
&:=& \sum_{k\ne0}\frac1kD_{\mu,k}D_{\nu,k}, \nonumber 
\end{eqnarray}
where 
\begin{equation}
\Delta_{(\pm, \pm, \pm)} \equiv 
\left( \pm \delta_{2l,m}\delta_{l,k+n} \pm \delta_{2k,m}\delta_{k,l+n} \pm \delta_{2n,m}\delta_{n,k+l} \right).\nonumber
\end{equation}

Then the results are given as follows:
\begin{eqnarray}
\left( \Pi_TX\Pi_Y,\Pi_TX\Pi_Y \right) 
&=& -\frac{(\alpha')^2}{2r^2}p^2D^3_{XY}+\frac{\alpha'}{8r^2}D^{(-,-,-)}_{TXY}, 
\nonumber \\
\left( \Pi_TY\Pi_X,\Pi_TY\Pi_X \right) 
&=& -\frac{(\alpha')^2}{2r^2}p^2D^3_{YX}+\frac{\alpha'}{8r^2}D^{(-,-,-)}_{TYX}, 
\nonumber \\
\left( \Pi_TX\Pi_Y,-\Pi_TY\Pi_X \right) 
&=& \frac{(\alpha')^2}{2r^2}p^2D^3_{XY}+\frac{\alpha'}{8r^2}\tilde{D}^{(+,+,+)}_{TXY}, 
\nonumber \\
\left( -\Pi_TY\Pi_X,\Pi_TX\Pi_Y \right) 
&=& \frac{(\alpha')^2}{2r^2}p^2D^3_{YX}+\frac{\alpha'}{8r^2}\tilde{D}^{(+,+,+)}_{TYX}, 
\nonumber \\
\left( \partial_\sigma TX\partial_\sigma Y,\partial_\sigma TX\partial_\sigma Y \right)
&=& \frac{\alpha'}{8r^2}D^{(-,-,-)}_{TXY}, 
\nonumber 
\end{eqnarray}

\begin{eqnarray}
\left( \partial_\sigma TY\partial_\sigma X,\partial_\sigma TY\partial_\sigma X \right) 
&=& \frac{\alpha'}{8r^2}D^{(-,-,-)}_{TYX}, 
\nonumber \\
\left( \partial_\sigma TX\partial_\sigma Y,-\partial_\sigma TY\partial_\sigma X \right) 
&=& \frac{\alpha'}{8r^2}\tilde{D}^{(-,-,+)}_{TXY}, 
\nonumber \\
\left( -\partial_\sigma TY\partial_\sigma X,\partial_\sigma TX\partial_\sigma Y \right) 
&=& \frac{\alpha'}{8r^2}\tilde{D}^{(-,-,+)}_{TYX}, 
\nonumber \\
\left( \Pi_TX\Pi_Y,\partial_\sigma TX\partial_\sigma Y \right) 
&=& \frac{\alpha'}{8r^2}D^{(-,+,-)}_{TXY}, 
\nonumber \\
\left( \Pi_TY\Pi_X,\partial_\sigma TY\partial_\sigma X \right) 
&=& \frac{\alpha'}{8r^2}D^{(-,+,-)}_{TYX}, 
\nonumber \\
(\Pi_TX\Pi_Y,-\partial_\sigma TY\partial_\sigma X) 
&=& \frac{\alpha'}{8r^2}\tilde{D}^{(-,+,+)}_{TXY},
\nonumber \\
(-\Pi_TY\Pi_X,\partial_\sigma TX\partial_\sigma Y) 
&=& \frac{\alpha'}{8r^2}\tilde{D}^{(-,+,+)}_{TYX}, 
\nonumber \\
\left( \partial_\sigma TX\partial_\sigma Y,\Pi_TX\Pi_Y \right) 
&=& \frac{\alpha'}{8r^2}D^{(-,+,-)}_{TXY}, 
\nonumber \\
\left( \partial_\sigma TY\partial_\sigma X,\Pi_TY\Pi_X \right) 
&=& \frac{\alpha'}{8r^2}D^{(-,+,-)}_{TYX}, 
\nonumber \\
(-\partial_\sigma TY\partial_\sigma X,\Pi_TX\Pi_Y) 
&=& \frac{\alpha'}{8r^2}\tilde{D}^{(-,+,+)}_{TXY}, 
\nonumber \\
(-\partial_\sigma TX\partial_\sigma Y,\Pi_TY\Pi_X) 
&=& \frac{\alpha'}{8r^2}\tilde{D}^{(-,+,+)}_{TYX}, 
\nonumber \\
(\partial_\sigma TY,\partial_\sigma TY) 
&=& -\frac2{\pi^2}D^3_{TY}. 
\nonumber 
\end{eqnarray}
It turned out that most of the terms cancel miraculously. 
As a result, we find 
\begin{eqnarray}
-\sum_{m\ne0}\frac1m[V_{1,-m}V_{1,m}]_D
&=& -\frac{\alpha'}{4r^2}\sum_{\stackrel{\scriptstyle n,k\ne0}{n+k\ne0}}\frac{n}{k}\Bigl[D_{T,n}D_{X,k}D_{Y,n+k}+D_{T,n+k}D_{X,k}D_{Y,n}
\nonumber \\
& & \hspace{2.3cm}+D_{T,n}D_{Y,k}D_{X,n+k}+D_{T,n+k}D_{Y,k}D_{X,n} \Bigr]
\nonumber \\
& & +\frac{\alpha'}{2r^2}\sum_{\stackrel{\scriptstyle k,l\ne0}{k+l\ne0}}D_{T,k+l}D_{X,k}D_{Y,l}-\frac{2}{\pi^2}\sum_{n\ne0}\frac{1}{n}D_{T,n}D_{Y,n}. \nonumber \\
\end{eqnarray}
To extract divergent terms, we use the following formulae: 
\begin{eqnarray}
\sum_{n\ne0}\frac1nD_{T,n}D_{Y,n} 
&=&N_T(2)+N_Y(2)+\zeta(1), \\
\sum_{\stackrel{\scriptstyle n,k\ne0}{n+k\ne0}}\frac nkD_{T,n}D_{X,k}D_{Y,n+k} 
&=& \zeta(1)\left( 2N_T(0)  + \zeta(-1) 
 \right)
 \nonumber \\ [-6mm]
& & +2N_X(2)N_Y(0)+2N_X(2)N_T(0)-N_{TX}(2) \nonumber \\
& & -\frac16N_X(2)-\frac12N_X(1)-\frac12N_X(0)-N_Y(0) \nonumber \\
& & +\sum_{n=1}^\infty\sum_{k=1}^n\frac1k\left( N_{Y,n}-N_{T,n} \right)\nonumber \\
&&+\sum_{k=1}^\infty\sum_{n\ne0,-k}\frac1{k(n+k)}N_{T,n}N_{Y,n+k},  \\ 
\sum_{\stackrel{\scriptstyle k,l\ne0}{k+l\ne0}}D_{T,k+l}D_{X,k}D_{Y,l} 
&=& N_{XY}(2)-N_{TX}(2)-N_{TY}(2)+N_X(0)+N_Y(0) \nonumber \\[-6mm]
& & -N_T(1)+N_T(0)  +\frac12\zeta(0)^2 
\end{eqnarray}
and so on. 
Using these results and the  following identity,
\begin{eqnarray}
\sum_{k=1}^\infty\sum_{n\ne0,-k}\frac1{k(n+k)}N_{T,n}N_{Y,n+k}+\sum_{k=1}^\infty\sum_{n\ne0,-k}\frac1{k(n+k)}N_{Y,n}N_{T,n+k} &=& N_{TY}(2), \nonumber 
\end{eqnarray}
we obtain the result of (\ref{V1V1-result}).  

\section{Useful formulae for traces and derivation of the partition function (\ref{partition function final})}
\label{App-trace-NXY}
In this appendix, we give general formulae for the traces such as 
\begin{equation}
{\rm tr} [ e^{-2\pi s {\tilde H}_0}  N_\mu(x)], \hspace{10mm} 
{\rm tr} [ e^{-2\pi s {\tilde H}_0}  N_{\mu \nu} (x)]
\end{equation}
where  ${\tilde H}_0$, $N_\mu(x)$,  $N_{\mu \nu} (x)$ are defined in 
in  Eq.(\ref{tildeH_0}) and Eq.(\ref{NalphaDef}). 
Using them, we derive the expression (\ref{partition function final}) for the partition function. 

The calculation of these traces is reduced to considering a single harmonic oscillator 
\begin{equation}
[\alpha_n,\alpha_{-n}]\ =\ n, 
\end{equation}
for which we determine 
\begin{equation}
{\rm tr}_n\left[ e^{-2\pi sN_n}N_n \right], \hspace{1cm} N_n\ :=\ \alpha_{-n}\alpha_n. 
\end{equation}
The trace ${\rm tr}_n$ is taken over the Fock space of $\alpha_n$. 
It is easy to obtain 
\begin{equation}
{\rm tr}_n\left[ e^{-2\pi sN_n}N_n \right]\ =\ \frac{nq^n}{1-q^n}\cdot\frac1{1-q^n}, 
\end{equation}
where $q=^{-2\pi s}$. 
Using this formulae, we obtain 
\begin{eqnarray}
{\rm tr}\left[ e^{-2\pi s\tilde{H}_0}N_\mu(x) \right] 
&=& \sum_{n=1}^\infty\frac{n^{1-x}q^n}{1-q^n}\cdot\prod_{m=1}^\infty(1-q^m)^{-3}, \\
{\rm tr}\left[ e^{-2\pi s\tilde{H}_0}N_{\mu\nu}(x) \right] 
&=& \sum_{n=1}^\infty\frac{n^{2-x}q^{2n}}{(1-q^n)^2}\cdot\prod_{m=1}^\infty(1-q^m)^{-3}, 
\end{eqnarray}
where we assumed $\mu\ne\nu$. 

These formulae are sufficient to determine the partition function (\ref{F}). 
For this purpose, it is helpful to notice the following formulae 
\begin{equation}
{\rm tr}\left[ e^{-2\pi s\tilde{H}_0}\left( 2N_{\mu\nu}(2)+N_\mu(1)+N_\nu(1) \right) \right]\ =\ \sum_{n=1}^\infty\frac{2q^n}{(1-q^n)^2}\cdot\prod_{m=1}^\infty(1-q^m)^{-3}. 
\end{equation}
Then we obtain 
\begin{eqnarray}
s\,{\rm tr}\left[ e^{-2\pi s\left( \tilde{H}_0-\frac18 \right)}H_2 \right] 
&=& s\left[ \frac{\alpha'}{r^2}\sum_{n=1}^\infty\frac{2q^n}{(1-q^n)^2}-\frac4{\pi^2}\sum_{n=1}^\infty\frac{n^{-1}q^n}{1-q^n} \right. \nonumber \\ [2mm] 
& & \left. \hspace{5mm} - \frac13 
\alpha'k^2 
+\epsilon_0 \right]\eta(is)^{-3},     \\ [1mm]  
\label{traceH2app}
-\pi s^2{\rm tr}\left[ e^{-2\pi s\tilde{H}_0}H_1^2 \right] 
&=& -\pi s^2\left( \frac{4(\alpha')^2}{r^2}k^2+\frac4{\pi^2} \right)\sum_{n=1}^\infty\frac{2q^n}{(1-q^n)^2}\eta(is)^{-3}. \nonumber \\ \label{traceH12app}
\end{eqnarray}
Interestingly, after performing the $k$-integration, 
one more cancellation occurs between these two traces; 
i.e., the momentum integration of  (\ref{traceH12app}) becomes
\begin{equation}
\int\frac{dk}{2\pi}e^{-2\pi \alpha'sk^2}{\rm tr}\left[ e^{-2\pi s\tilde{H}_0}H_1^2 \right]\ =\ (8\pi^2\alpha's)^{-\frac12}\left( -\frac{\alpha'}{r^2}s-\frac4\pi s^2 \right)\sum_{n=1}^\infty\frac{2q^n}{(1-q^n)^2}\eta(is)^{-3}
\end{equation}
whose first term in the parenthesis is cancelled by the momentum integration
of the first term in the square bracket of (\ref{traceH2app}), 
\begin{equation}
\int\frac{dk}{2\pi}e^{-2\pi \alpha'sk^2}{\rm tr}\left[ e^{-2\pi s\tilde{H}_0}H_2 \right]\ =\ (8\pi^2\alpha's)^{-\frac12}\cdot\frac{\alpha'}{r^2}s\sum_{n=1}^\infty\frac{2q^n}{(1-q^n)^2}\eta(is)^{-3}+\cdots. 
\end{equation}

Finally, the partition function becomes 
\begin{eqnarray}
& & \int_0^\infty\frac{ds}{2s}(8\pi^2\alpha's)^{-\frac12}e^{-\frac{2r^2}{\pi\alpha'}s}\eta(is)^{-24}\left( 1-
 \frac13  v^2 \right)^{-\frac12} \nonumber \\
& & \times\left[ 1-2\pi v^2\left( -\frac4{\pi^2}s\sum_{n=1}^\infty\frac{n^{-1}q^n}{1-q^n}+\epsilon_0s-\frac4{\pi}s^2\sum_{n=1}^\infty\frac{2q^n}{(1-q^n)^2} \right) \right]+{\cal O}(v^4). \nonumber \\
\end{eqnarray}

\section{Matrix elements for low lying states}
\label{sec-matrixelement}
We determine the matrix elements $\langle\vec{\bm n};k|O|\vec{\bm n}';k\rangle$ for $O=H_1^2,H_2$, in order to calculate the eigenvalues $E_i(k,r,\omega)$ of $H_0(v)$ we need in subsection \ref{sec-diagonalization}. 
The states $|\vec{\bm n}\rangle$ we consider in the following satisfy $N(\vec{\bm n})\le1$. 

First for the ground state with $N(\vec{\bm n})=0$, all of $n^{\alpha}_{n}$ are zero. 
Therefore, we obtain 
\begin{eqnarray}
\langle\vec{\bm n};k|H_1^2|\vec{\bm n};k\rangle 
&=& 0, \\
\langle\vec{\bm n};k|H_2|\vec{\bm n};k\rangle 
&=& -\frac13\alpha'k^2+\epsilon_0. 
\label{matrixelement0}
\end{eqnarray}
 After returning to the Lorentzian metric, $-v^2 H_2/\alpha'$ gives the $v^2$ corrections
to the energy of the tachyonic state. 

Then we consider the first excited states  with $N(\vec{\bm n})=1$,
{\color{black} which correspond to the  massless open string states
excited by world sheet variables $T, X, Y$}; so there are three states specified by the integers $(n_1^T,n_1^X,n_1^Y)$. 
We denote these states by 
\begin{equation}
\vec{\bm n}_1\ =\ (1,0,0), \hspace{5mm} \vec{\bm n}_2\ =\ (0,1,0), \hspace{5mm} \vec{\bm n}_3\ =\ (0,0,1). 
\end{equation}
The diagonal matrix elements can be obtained easily. 
The results are 
\begin{eqnarray}
\langle\vec{\bf n}_1;k|H_2|\vec{\bf n}_1;k\rangle 
&=& -\frac13\alpha'k^2-\frac4{\pi^2}+\epsilon_0, 
\label{matrixelement11} \\
\langle\vec{\bf n}_2;k|H_2|\vec{\bf n}_2;k\rangle 
&=& \frac{\alpha'}{r^2}-\frac13\alpha'k^2+\epsilon_0, 
\label{matrixelement12} \\
\langle\vec{\bf n}_3;k|H_2|\vec{\bf n}_3;k\rangle 
&=&  \frac{\alpha'}{r^2}-\frac13\alpha'k^2+\epsilon_0. 
\label{matrixelement13}
\end{eqnarray}

We also need the off-diagonal matrix elements. 
To obtain them, we must use $H_1$ and $H_2$, not just their diagonal parts $[H_1^2]_D$ and $[H_2]_D$. 
Using the expression (\ref{H_1}), we find that the only non-zero off-diagonal elements of $H_1$ are 
\begin{equation}
\langle\vec{\bm n}_1;k|H_1|\vec{\bm n}_3;k\rangle\ =\ \frac{2i}{\pi}, \hspace{1cm} \langle\vec{\bm n}_2;k|H_1|\vec{\bm n}_3;k\rangle\ =\ \frac{2i\alpha'}{r}k, 
\end{equation}
and their complex conjugates. 

It turns out that most of terms in $H_2$ do not contribute to the off-diagonal elements for the states with $N(\vec{\bm n})=1$. 
This can be seen by considering the following matrix element 
\begin{equation}
\langle0|\alpha_m\alpha_n|1\rangle, \hspace{1cm} |1\rangle\ :=\ |n_1=1\rangle. 
\end{equation}
The matrix elements for $(m,n)=(1,0),(0,1)$ are non-vanishing only if $\langle0|\alpha_0|0\rangle$ is non-zero, and the other matrix elements vanish. 
This implies that terms with a zero mode in $H_2$ give non-zero off-diagonal matrix elements. 

Among the worldsheet fields, $\Pi_T$ is the only field which has a zero mode. 
It turns out that 
\begin{equation}
-\int_0^\pi d\sigma\,\frac{2\pi\alpha'}{r^2}x(\sigma)\Pi_T^2X. 
\end{equation}
is the only term in $V_2$ which gives the off-diagonal matrix element 
\begin{eqnarray}
\langle\vec{\bm n}_1;k|\left[ -\int_0^\pi d\sigma\,\frac{2\pi\alpha'}{r^2}x(\sigma)\Pi_T^2X \right]|\vec{\bm n}_2;k\rangle 
&=& -\frac{2\alpha'}{\pi r}k. 
\end{eqnarray}
and its complex conjugate.

The terms in $-\sum_{m\ne0}V_{1,-m}V_{1,m}/m$ which could possibly give off-diagonal matrix elements are 
\begin{eqnarray}
-\sum_{m\ne0}\frac1m\int_0^\pi d\sigma\frac{-2\pi\alpha'}{r^2}\left[ \Pi_T\left( X\Pi_Y-Y\Pi_X \right) \right]_{-m}\int_0^\pi d\sigma'\frac{2r^2}{\pi^2\alpha'}\left[ \partial_\sigma TY \right]_m \nonumber \\
\end{eqnarray}
and its Hermitian conjugate. 
In fact, we find that the off-diagonal matrix elements of these terms vanish. 

In summary, the matrix elements of $H_0(v)$ for the states with $N(\vec{\bm n})\le1$ are 
\begin{equation}
\left( 1-\frac13v^2 \right)\alpha'k^2+\frac{r^2}{\pi^2\alpha'}+v^2\epsilon_0+\left[
\begin{array}{cccc}
0 & 0 & 0 & 0 \\ [2mm]
0 & \displaystyle{1-\frac4{\pi^2}v^2} & \displaystyle{-\frac{2\alpha'}{\pi r}kv^2} & \displaystyle{\frac{2i}{\pi}v} \\ [4mm]
0 & \displaystyle{-\frac{2\alpha'}{\pi r}kv^2} & \displaystyle{1+\frac{\alpha'}{r^2}v^2} & \displaystyle{\frac{2i\alpha'}{r}kv} \\ [4mm]
0 & \displaystyle{-\frac{2i}{\pi}v} & \displaystyle{-\frac{2i\alpha'}{r}kv} & \displaystyle{1+\frac{\alpha'}{r^2}v^2}
\end{array}
\right]. 
\end{equation}

 Back to the Lorentzian metric and multiplying $-v^2/\alpha'$,  
these expressions give $v^2$-dependent corrections to the mass matrix 
of the open string massless states, whose 0-th order energy is given by
$k^2 + r^2/\pi^2 \alpha^{'2}$. Since $H_1$ and $H_2$ mix these 3 states, 
we need to diagonalize the matrix to obtain the energy eigenvalues. 
It is given in subsection \ref{sec-diagonalization}.


\section{Integration for ${\cal V}_{02}$ and ${\cal V}_{03}$} \label{massless V}

Recall that $E_2(k,r,\omega)-1$ is given as 
\begin{eqnarray}
E_2(k,r,\omega)-1 
&=& \left( 1-\frac13\omega^2r^2 \right)\alpha'k^2+(1+\epsilon_0\pi^2\alpha'\omega^2)\frac{r^2}{\pi^2\alpha'} \nonumber \\
& & +2\sqrt{\alpha'}\omega h(k,r)^\frac12+\alpha'\omega^2-2\alpha'\omega^2\frac{r^2}{\pi^2\alpha'}-\frac{\omega^2r^2}{2\pi^2}h(k,r)^{-1}. \nonumber \\
\end{eqnarray}
The integral 
\begin{equation}
{\cal V}_{02}\ =\ -\int_0^\infty\frac{ds}{2s}\int_{-\infty}^{+\infty}\frac{dk}{2\pi}\,e^{-2\pi s(E_2(k,r,\omega)-1)}
\end{equation}
is very complicated. 
To see the behavior of ${\cal V}_{02}$, we replace 
\begin{eqnarray}
E_2(k,r,\omega)-1 
&\to& \left( 1-\frac13\omega^2r^2 \right)\alpha'k^2+(1+\epsilon_0\pi^2\alpha'\omega^2)\frac{r^2}{\pi^2\alpha'} \nonumber \\
& & +2{\alpha'}\omega |k|+\alpha'\omega^2-2\alpha'\omega^2\frac{r^2}{\pi^2\alpha'}. \nonumber \\
\end{eqnarray}
Then, the integral becomes 
\begin{eqnarray}
{\cal V}_{02} 
&\to& -2\int_0^\infty\frac{ds}{2s}\,e^{-2\pi s\left[ \alpha'\omega^2+(1+(\epsilon_0\pi^2-2)\alpha'\omega^2)\frac{r^2}{\pi^2\alpha'} \right]} \nonumber \\
& & \times\int_0^\infty\frac{dk}{2\pi}e^{-2\pi s\alpha'\left[ \left( 1-\frac13\omega^2r^2 \right)k^2+2\omega k \right]} .
\end{eqnarray}
The expression for ${\cal V}_{03}$ is obtained by flipping the sign of $\omega$. 
It can be written as 
\begin{eqnarray}
{\cal V}_{03} 
&\to& -2\int_0^\infty\frac{ds}{2s}\,e^{-2\pi s\left[ \alpha'\omega^2+(1+(\epsilon_0\pi^2-2)\alpha'\omega^2)\frac{r^2}{\pi^2\alpha'} \right]} \nonumber \\
& & \times\int_{-\infty}^0\frac{dk}{2\pi}e^{-2\pi s\alpha'\left[ \left( 1-\frac13\omega^2r^2 \right)k^2+2\omega k \right]} .
\end{eqnarray}
The sum of them can be easily integrated and results in (\ref{EP-massless3}).

\section{$[H_2^{(l)}]_D$ for the linear system} \label{linear H_2}

$[H_2^{(l)}]_D$ for the linear system can be calculated in a similar way as the revolving system. 
The difference is that $H_0^{(l)}$ involves the term proportional to $t^2$ and the zero mode forms a harmonic oscillator. 
In terms of the creation/annihilation operators of the harmonic oscillator, 
$t$ and $p$ are written as
\begin{equation}
t= \sqrt{\frac{\pi \alpha'}{2v}}(\alpha_0+\alpha_0^\dagger), \hspace{1cm}
p= -i\sqrt{\frac{v}{2\pi \alpha'}}(\alpha_0-\alpha_0^\dagger).
\end{equation}
Then the free Hamiltonian for the zero modes in $H_0^{(l)}$ is
given by $\frac{2v}\pi (N_0+\frac12)$,
where $[\alpha_0,\alpha_0^\dagger]=1$ and $N_0:=\alpha_0^\dagger \alpha_0$.

For products including $\tilde T$ and $\Pi_T$, the diagonal parts are given as 
\begin{eqnarray}
\left[ \Pi_T(\sigma)\Pi_T(\sigma') \right]_D 
&=& \frac{vr^2}{\pi^3 \alpha'}\left(N_0+\frac12\right)+\frac{r^2}{2\pi^2\alpha'}\sum_{n\ne0}nD_{T,n}\cos n\sigma\cos n\sigma', \nonumber \\
\left[ \tilde T(\sigma) \tilde T(\sigma') \right]_D 
&=& \frac{\pi \alpha'}{vr^2}\left(N_0+\frac12\right)+\frac{2\alpha'}{r^2}\sum_{n\ne0}\frac1nD_{T,n}\cos n\sigma\cos n\sigma', \nonumber \\
\left[ \partial_\sigma T(\sigma) \partial_\sigma T(\sigma') \right]_D 
&=& \frac{2\alpha'}{r^2}\sum_{n\ne0}nD_{T,n}\sin n\sigma\sin n\sigma', \nonumber \\
\left[ \Pi_T(\sigma)\tilde T(\sigma') \right]_D 
&=& -\frac{i}{2\pi}-\frac i\pi\sum_{n\ne0} D_{T,n}\cos n\sigma\cos n\sigma', \nonumber \\
\left[ \tilde T(\sigma)\Pi_T(\sigma') \right]_D 
&=& \frac{i}{2\pi}+\frac i\pi\sum_{n\ne0} D_{T,n}\cos n\sigma\cos n\sigma'. \nonumber 
\end{eqnarray}

For $X$, $\Pi_X$ and $Y$, $\Pi_Y$, mode expansions are given in the same form as
those in the revolving system.

\vspace{5mm}
Let us first look at differences between the revolving and the linear systems. 
Since $T$ and $X$ obey different, Neumann and Dirichlet,  boundary conditions, 
these terms give  different contributions to the partition function. 
For example, for the revolving system, $[H_2]$ has a contribution 
\begin{eqnarray}
\int_0^\pi d\sigma\,\frac{\pi\alpha'}{r^2}[X^2][\Pi_Y^2] 
&=& \frac{\alpha'}{4r^2}\Biggl[ \zeta(1) \Bigl( 2N_Y(0)-\frac1{12} \Bigr)+4N_X(2)N_Y(0)-\frac16N_X(2) \Biggr] \nonumber \\
& & +\frac{\alpha'}{4r^2}\Biggl[ 2N_{XY}(2)+N_X(1)+N_Y(1)-\frac12-\frac12\zeta(0) \Biggr], \nonumber 
\end{eqnarray}
while for the linear system, the corresponding contribution to $[H^{(l)}_2]$ is 
\begin{eqnarray}
\int_0^\pi d\sigma\,\frac{\pi\alpha'}{r^2}[T^2][\Pi_Y^2] 
&=& \frac{\alpha'}{4r^2}\Biggl[ \zeta(1) \Bigl( 2N_Y(0)-\frac1{12} \Bigr)+4N_X(2)N_Y(0)-\frac16N_X(2) \Biggr] \nonumber \\
& & -\frac{\alpha'}{4r^2}\Biggl[ 2N_{XY}(2)+N_X(1)+N_Y(1)-\frac12-\frac12\zeta(0) \Biggr],\nonumber 
\end{eqnarray}
in which the second line has the opposite sign. 
Note that the divergent terms are common in two cases. 
This turns out to be the case for all divergent terms. 
\vspace{5mm}

In a way similar to the revolving system, each terms contributing to $[V_2^{(l)}]_D$ are calculated as follows:
\begin{eqnarray}
\int_0^\pi d\sigma\,\frac{\pi\alpha'}{r^2}[\tilde T^2]_D[\Pi_Y^2]_D 
&=& \frac{\alpha'}{4r^2}\left( D^1_{TY}-D^2_{TY}+\frac12D_T(0) \right) \nonumber \\
& & +\frac{\pi \alpha'}{4vr^2}\left(N_0+\frac12\right)D_Y(-1), 
\nonumber \\
\int_0^\pi d\sigma\,\frac{\pi\alpha'}{r^2}[Y^2]_D[\Pi_T^2]_D 
&=& \frac{\alpha'}{4r^2}\left( D^1_{YT}-D^2_{YT}+\frac12D_Y(0) \right) \nonumber \\
& & +\frac{v\alpha'}{\pi r^2}\left(N_0+\frac12\right)D_Y(1), 
\nonumber \\
-\int_0^\pi d\sigma\,\frac{\pi\alpha'}{r^2}[\tilde T\Pi_T]_D[\Pi_YY]_D 
&=& 0, \nonumber \\
-\int_0^\pi d\sigma\,\frac{\pi\alpha'}{r^2}[\Pi_YY]_D[\tilde T\Pi_T]_D 
&=& 0, \nonumber 
\end{eqnarray}
\begin{eqnarray}
\int_0^\pi d\sigma\,\frac{r^2}{4\pi\alpha'}[(\partial_\sigma X)^2]_D[\tilde T^2]_D 
&=& \frac{\alpha'}{4r^2}\left( D^1_{XT}+D^2_{XT}-\frac12D_T(0) \right) \nonumber \\
& & +\frac{\pi \alpha'}{4vr^2}\left(N_0+\frac12\right)D_X(-1),
\nonumber \\
\int_0^\pi d\sigma\,\frac{r^2}{4\pi\alpha'}[(\partial_\sigma X)^2]_D[Y^2]_D 
&=& \frac{\alpha'}{4r^2}\left( D^1_{XY}-D^2_{XY}+\frac12D_Y(0) \right),
\nonumber \\
\int_0^\pi d\sigma\,\frac{r^2}{4\pi\alpha'}\frac{4}{\pi^2}[T^2]_D 
&=& \frac1{\pi^2}D_T(1),
\nonumber \\
\int_0^\pi d\sigma\,\frac{r^2}{4\pi\alpha'}\frac{4}{\pi^2}[Y^2]_D 
&=& \frac1{\pi^2}D_Y(1). 
\nonumber 
\end{eqnarray}

Summing all of them, and using the same formulae as the revolving system, we get
\begin{eqnarray}
\left[V_2\right]_D
&=&\zeta(1)\left[\frac{2}{\pi^2}+\frac{\alpha'}{2r^2}\left(N_T(0)+2N_X(0)+N_Y(0)-\frac16\right)\right]\nonumber \\
&&+\frac{\alpha'}{r^2}\Bigl[-N_{TY}(2)+\frac12N_{TX}(2)-\frac12N_{XY}(2)\nonumber \\[2mm]
&&\hspace{1cm}+N_T(0)N_Y(2)+N_T(2)N_Y(0)+N_T(2)N_X(0)+N_X(0)N_Y(2)\nonumber \\[2mm]
&&\hspace{1cm}-\frac1{12}N_T(2)-\frac1{12}N_Y(2)-\frac14N_T(1)-\frac34N_Y(1)+\frac18\Bigr]\nonumber \\[2mm]
&&+\frac{\pi \alpha'}{2vr^2}\left(N_0+\frac12\right)\left(N_X(0)+N_Y(0)-\frac1{12}\right)+\frac{2}{\pi^2}\bigl(N_T(2)+N_Y(2)\bigr)\nonumber \\
&&+\mathcal{O}(v). \label{linear V_2-D}
\end{eqnarray}

We need to take some care when calculating the contribution to $[V_1^{(l)}V_1^{(l)}]_D$ that includes zero modes. For example, there is a following term in $(\Pi_X \tilde T\Pi_Y, \Pi_X \tilde T\Pi_Y)$:
\begin{eqnarray}
&&-\frac{\pi \alpha'}{2vr^2}\sum_{m=-\infty}^\infty\sum_{n,k\neq0}\left(\frac{\alpha_0\alpha_0^\dagger}{2v/\pi+m}+\frac{\alpha_0^\dagger\alpha_0}{-2v/\pi+m}\right)\nonumber \\
&&\hspace{3cm}\times nkD_{X,n}D_{Y,k}\delta_{m,n+k}\left[\frac1\pi \int d\sigma \sin n\sigma \sin k\sigma \right]^2\nonumber \\
&=&-\frac{\pi^2 \alpha'}{16v^2r^2}\sum_{n\neq0}n^2\bigl(D_{X,n}D_{Y,n}-D_{X,n}\bigr)\nonumber \\
&&-\frac{\pi \alpha'}{8vr^2}\sum_{n\neq0}\frac{k^2}{2v/\pi+2n}\Bigl[D_{X,}D_{Y,n}+N_0\bigl(D_{X,n}+D_{Y,n}-1\bigr)\Bigr]. \nonumber 
\end{eqnarray}

The terms includes no zero modes can be calculated in the same way as the revolving system. Thus each contributions are obtained as follows:
\begin{eqnarray}
\left( \Pi_X \tilde T \Pi_Y,\Pi_X \tilde T \Pi_Y\right) 
&=& -\frac{\alpha'}{8r^2}D^{(+,+,+)}_{XTY}-\frac{\pi^2 \alpha'}{16v^2r^2}\sum_{n\neq0}n^2\bigl(D_{X,n}D_{Y,n}-D_{X,n}\bigr)\nonumber \\
&&\hspace{-5mm}-\frac{\pi \alpha'}{8vr^2}\sum_{n\neq0}n^2\frac{D_{X,n}D_{Y,n}+N_0\bigl(D_{X,n}+D_{Y,n}-1\bigr)}{2v/\pi+2n},\nonumber \\
\left( -\Pi_X\Pi_T Y,-\Pi_X\Pi_T Y \right) 
&=&-\frac{\alpha'}{8r^2}D^{(+,+,+)}_{XYT}
-\frac{\alpha'}{4r^2}\sum_{n\neq0}\bigl(D_{X,n}D_{Y,n}-D_{X,n}\bigr)\nonumber \\
&& \hspace{-5mm} -\frac{v \alpha'}{2\pi r^2}\sum_{n\neq0}\frac{D_{X,n}D_{Y,n}+N_0\bigl(D_{X,n}+D_{Y,n}-1\bigr)}{2v/\pi+2n},\nonumber 
\end{eqnarray}
\begin{eqnarray}
\left( \partial_\sigma X \tilde T \partial_\sigma Y,\partial_\sigma X \tilde T \partial_\sigma Y \right) 
&=& -\frac{\alpha'}{8r^2}D^{(+,+,+)}_{XTY}
-\frac{\pi^2 \alpha'}{16v^2r^2}\sum_{n\neq0}n^2\bigl(D_{X,n}D_{Y,n}-D_{X,n}\bigr) \nonumber \\
&&\hspace{-5mm} -\frac{\pi \alpha'}{8vr^2}\sum_{n\neq0}n^2\frac{D_{X,n}D_{Y,n}+N_0\bigl(D_{X,n}+D_{Y,n}-1\bigr)}{2v/\pi+2n},\nonumber \\
\left( -\partial_\sigma X \partial_\sigma T Y,-\partial_\sigma X \partial_\sigma T Y \right) 
&=& -\frac{\alpha'}{8r^2}D^{(+,+,+)}_{XYT},
\nonumber \\
\left( \partial_\sigma T Y,\partial_\sigma T Y \right) 
&=& -\frac2{\pi^2}D^3_{TY},
\nonumber
\end{eqnarray}

\begin{eqnarray}
\left( \Pi_X \tilde T \Pi_Y,-\Pi_X\Pi_T Y \right) 
&=& \left( -\Pi_X\Pi_T Y,\Pi_X \tilde T \Pi_Y \right)\nonumber \\
&=& \frac{\alpha'}{8r^2}\tilde{D}^{(+,+,+)}_{XTY}
-\frac{\pi \alpha'}{8vr^2}(2N_0+1)\sum_{n\neq0}n\bigl(D_{X,n}D_{Y,n}-D_{X,n}\bigr)\nonumber \\
&&\hspace{-5mm} +\frac{\alpha'}{4r^2}\sum_{n\neq0}n\frac{D_{X,n}D_{Y,n}+N_0\bigl(D_{X,n}+D_{Y,n}-1\bigr)}{2v/\pi+2n},\nonumber \\
\left( \Pi_X \tilde T \Pi_Y,\partial_\sigma X\tilde T \partial_\sigma Y \right) 
&=& \left( \partial_\sigma X\tilde T \partial_\sigma Y, \Pi_X \tilde T \Pi_Y \right)\nonumber \\
&=& -\frac{\alpha'}{8r^2}D^{(+,-,+)}_{XTY}
+\frac{\pi^2 \alpha'}{16v^2r^2}\sum_{n\neq0}n^2\bigl(D_{X,n}D_{Y,n}-D_{X,n}\bigr)\nonumber \\
&&\hspace{-5mm} -\frac{\pi \alpha'}{8vr^2}\sum_{n\neq0}n^2\frac{D_{X,n}D_{Y,n}+N_0\bigl(D_{X,n}+D_{Y,n}-1\bigr)}{2v/\pi+2n},\nonumber \\
\left( \Pi_X \tilde T \Pi_Y,-\partial_\sigma X \partial_\sigma T Y \right) 
&=& \left( -\partial_\sigma X \partial_\sigma T Y, \Pi_X \tilde T \Pi_Y \right)\nonumber \\
&=& -\frac{\alpha'}{8r^2}\tilde{D}^{(+,-,-)}_{XTY},
\nonumber \\
\left( -\Pi_X\Pi_T Y,\partial_\sigma X\tilde T \partial_\sigma Y \right) 
&=& \left( \partial_\sigma X\tilde T \partial_\sigma Y, -\Pi_X\Pi_T Y \right)\nonumber \\
&=& \frac{\alpha'}{8r^2}\tilde{D}^{(+,-,+)}_{XTY}
+\frac{\pi \alpha'}{8vr^2}(2N_0+1)\sum_{n\neq0}n\bigl(D_{X,n}D_{Y,n}-D_{X,n}\bigr)\nonumber \\
&&\hspace{-5mm} +\frac{\alpha'}{4r^2}\sum_{n\neq0}n\frac{D_{X,n}D_{Y,n}+N_0\bigl(D_{X,n}+D_{Y,n}-1\bigr)}{2v/\pi+2n},\nonumber \\
\left( -\Pi_X\Pi_T Y,-\partial_\sigma X \partial_\sigma T Y \right) 
&=& \left( -\partial_\sigma X \partial_\sigma T Y, -\Pi_X\Pi_T Y \right)\nonumber \\
&=& \frac{\alpha'}{8r^2}D^{(-,+,-)}_{XYT},
\nonumber \\
\left( \partial_\sigma X\tilde T \partial_\sigma Y,-\partial_\sigma X \partial_\sigma T Y \right) 
&=& \left( -\partial_\sigma X \partial_\sigma T Y, \partial_\sigma X\tilde T \partial_\sigma Y \right)\nonumber \\
&=& -\frac{\alpha'}{8r^2}\tilde{D}^{(+,+,-)}_{XTY}.
\nonumber 
\end{eqnarray}

Again, most of the terms cancel. The remaining contributions are 
\begin{eqnarray}
&&-\sum_{m\neq0}\frac1m\left[V^{(l)}_{1,-m}V^{(l)}_{1,m}\right]_D\nonumber \\
&=&-\zeta(1)\Biggl[\frac{2}{\pi^2}+\frac{\alpha'}{2r^2}\Bigl\{N_T(0)+2N_X(0)+N_Y(0)-\frac16\Bigr\}\Biggr]\nonumber \\
&&-\frac{\alpha'}{r^2}\Biggl[-N_{TY}(2)+\frac12N_{TX}(2)-\frac12N_{XY}(2)\nonumber \\
&&\hspace{1cm}+N_T(0)N_Y(2)+N_T(2)N_Y(0)+N_T(2)N_X(0)+N_X(0)N_Y(2)\nonumber \\[2mm]
&&\hspace{1cm}-\frac1{12}N_T(2)-\frac1{12}N_Y(2)-\frac14N_T(1)-\frac34N_Y(1)\nonumber \\
&&\hspace{1cm}-N_T(0)-N_X(0)-N_Y(0)+\frac14\Biggr]\nonumber \\
&&-\frac{\pi \alpha'}{2vr^2}\left(N_0+\frac12\right)\left(N_X(0)+N_Y(0)-\frac1{12}\right)+\frac{2}{\pi^2}\bigl(N_T(2)+N_Y(2)\bigr)\nonumber \\[2mm]
&&+\mathcal{O}(v).\label{linear V1V1-D}
\end{eqnarray}

As a result, summing up (\ref{linear V_2-D}) and (\ref{linear V1V1-D}), we find 
\begin{equation}
[H_2^{(l)}]_D=\frac{\alpha'}{r^2}\left[N_T(0)+N_X(0)+N_Y(0)-\frac18\right] ,
\end{equation}
up to higher orders of $v$.


\end{document}